\documentstyle[preprint,aps,epsf,floats,eqsecnum]{revtex}

\input epsf

\tighten

\def\peta{\frac{g^2_A}{2 f^2}} 
\def\ddq{{{\rm d}^Dq \over (2\pi)^D}\,} 
\def\ddl{{{\rm d}^Dl \over (2\pi)^D}\,}
\def\ddqm{{{\rm d}^{D-1}{\bf q} \over (2\pi)^{D-1}}\,}
\def\ddlm{{{\rm d}^{D-1}{\bf l} \over (2\pi)^{D-1}}\,}
\def\ddkm{{{\rm d}^{D-1}{\bf k} \over (2\pi)^{D-1}}\,}
\def\dtqm{{{\rm d}^{3}{\bf q} \over (2\pi)^{3}}\,}
\def\dtlm{{{\rm d}^{3}{\bf l} \over (2\pi)^{3}}\,}
\def\dtkm{{{\rm d}^{3}{\bf k} \over (2\pi)^{3}}\,}
\def\bfq{{\bf q}} 
\def\bfk{{\bf k}} 
\def\bfp{{\bf p}}
\def\bfl{{\bf l}}
\def\bfpp{{\bf p '}} 
\def\darr#1{\raise1.5ex\hbox{$\leftrightarrow$}\mkern-16.5mu #1}
\def\){\right)} 
\def\({\left(} 
\def\]{\right]} 
\def\[{\left[}
\def\lskip{\vspace{\baselineskip}}

\newcommand{\eqn}[1]{\label{eq:#1}}
\newcommand{\refeq}[1]{(\ref{eq:#1})}

\newcommand{\Eq}{Eq.~\refeq} 

\newcommand{\beq}{\begin{eqnarray}}
\newcommand{\eeq}{\end{eqnarray}}

\newcommand{\mcal}[1]{{\mathcal #1}}
\newcommand{\makefig}[4]{\begin{figure}[t] 
                           \centerline{\epsfysize=#3 in \epsfbox{#2}} 
                           \caption{#4 \label{#1}} 
                         \end{figure}}
\newcommand{\maketwofigs}[6]{\begin{figure}[t]
                           \centerline{\epsfysize=#5 in \epsfbox{#2}
                                \hspace{#3 in} \epsfysize=#5 in \epsfbox{#4}}
                                \caption{#6 \label{#1}} 
                                \end{figure}}
 
\def\Journal#1#2#3#4{{#1} {\bf #2}, #3 (#4)}

\def\NPB{{\em Nucl. Phys.} B}
\def\NPA{{\em Nucl. Phys.} A}

\def\PLB{{\em Phys. Lett.} B}
\def\PRL{\em Phys. Rev. Lett.}

\def\PRC{{\em Phys. Rev.} C}

\def\PR{{\em Phys. Rev.} }

\def\FBS{{\em Few Body Systems Suppl.}}

\begin{document}
\preprint{\vbox{
\hbox{ NT@UW-99-15}
}}
\bigskip
\bigskip

\title{ NNLO Calculation of Two-Nucleon Scattering in an Effective Field Theory for a Two Yukawa Toy Model}
\author{Gautam Rupak and Noam Shoresh} 
\address{ Department of Physics, University of Washington, Seattle, WA 98195}
\maketitle

\begin{abstract}
The effective field theory (EFT) for a toy model of nucleons interacting via  a short range and a long range Yukawa potential is presented. The scattering amplitude in the ${}^1S_0$ channel is calculated up to next-to-next-to-leading order (NNLO) using the Kaplan-Savage-Wise (KSW) power counting scheme. A particular expansion of the amplitude about the pole at low (imaginary) momentum is used to derive matching conditions for the EFT couplings. In addition, after imposing  constraints from a Renormalization Group flow analysis, there are no new free parameter in the amplitude at NNLO. Comparing the NNLO phase shift to the full model we show that the EFT expansion is converging.
\end{abstract}

\begin{section}{Introduction}
\label{INTRO}
 
Effective field theory (EFT) can be a very useful tool in the study of
nuclear physics at low energies \cite{weinberg}-\cite{threebod}. Most importantly, it
provides a perturbative description of physical processes at momentum
$p$ well below a certain cutoff $\Lambda$. If fields for the lighter
degrees of freedom are included explicitly, and the effects of the
heavier particles are encoded in a derivative expansion of local
operators, one expects effects of the high-energy physics to be
suppressed by powers of $p/\Lambda$. Furthermore, a
perturbative expansion allows a systematic estimation of errors at any
order of the perturbation.

The contact terms in the Lagrangian that replace high-energy, short-range interactions, have dimensionful couplings. The mass scales that enter these couplings originate in the microscopic theory, and they are usually assumed to be ``natural'', i.e. of $\mcal{O}(\Lambda)$. For natural sized couplings, the derivative expansion of contact terms in the Lagrangian is an expansion in powers of $p/\Lambda$.

However, constructing an EFT for nucleons requires special care. The
reason is the appearance of other length scales, in particular the
${}^1S_0$ scattering length, $a$, which is much larger than the range
of the interactions ($a\sim 1/8\ \mathrm{MeV}\gg 1/m_{\pi}
, 1/m_{\rho}$). In this case the coupling constants need not
follow the natural scaling of $1/\Lambda$, but rather may
make up the dimensions by factors of $a$.

To understand the problem better, one first considers nucleons interacting through a single short range potential\cite{ksw1}. Lacking other ``light'' scales in the problem, one expects the scattering amplitude at low energies to be a simple Taylor expansion in $p/\Lambda$ of the exact amplitude. The scattering amplitude is related to the phase shift in the ${}^1S_0$ channel through
\beq
{\cal A}=\frac{4\pi}{M}\frac{1}{p\cot\delta-ip},
\eqn{pcot2A}
\eeq
where $p=\sqrt{ME}$ is the momentum of each nucleon in the
center of mass coordinates and $\delta$ is the ${}^1S_0$ phase shift.
The problem can be analyzed with the tools of quantum mechanics (which applies to the non-relativistic, heavy nucleons) where one writes what is known as the Effective Range Expansion (ERE):
\beq  
p\cot\delta=-\frac{1}{a}+\frac{1}{2}\sum_{n=0}^{\infty}r_{n}p^{2(n+1)}.
\eqn{ERE}
\eeq
The scattering length $a$ can be arbitrarily large, while the other length scales $r_{0},r_{1},\ldots$ are assumed to all have natural sizes, $r_n\sim1/\Lambda^{(2n+1)}$. $p$ is taken to be small compared to $\Lambda$, but not compared to $1/a\ll \Lambda$. 

Substituting \Eq{ERE} in \Eq{pcot2A}, it can be seen that there is a pole in the amplitude at $p\approx i/a$, which sets a small radius of convergence for the naive Taylor expansion about $p=0$. This explains the breakdown of the simple approach at momenta $p\sim 1/a$, and also suggests the solution - the EFT must be constructed in such a way as to include the pole in the amplitude at all orders.

In \cite{ksw1} the following expansion of the amplitude was presented: 
\beq
\mcal{A} &=& -{4\pi\over M}{1\over (1/a + i p)}\[ 1 + {r_0/2 \over (1/a + ip)}p^2 +  
{(r_0/2)^2\over (1/a + ip)^2} p^4 \right.\nonumber\\
& &\left.\hspace{2 in}+{(r_0/2)^3\over (1/a + ip)^3} p^6+ {(r_1/2)\over (1/a + ip) } p^4 +\ldots\].
\eqn{EREinAksw}
\eeq
In this paper we examine a different expansion \cite{bethe} in which the exact pole in the amplitude appears explicitly. We rewrite \refeq{ERE} as
\beq  
p\cot\delta=-\gamma+\frac{1}{2}\sum_{n=0}^{\infty}s_{n}\(p^2+\gamma^2\)^{n+1}.
\eqn{ERE2}
\eeq
where $i\gamma$ is the pole in the amplitude. 
We expand the amplitude in $p/\Lambda$ and $\gamma/\Lambda$. The coefficients $s_n$ and $r_n$ are the same at leading order (LO). Comparing the expansions Eqns.~\refeq{ERE} and \refeq{ERE2}, one can solve perturbatively for the relation between the coefficients $r_n$ and $s_n$. The first two terms are
\beq
\frac{1}{a}&=&\gamma -\frac{1}{2} {{\gamma }^2} {s_0}+\mcal{O}({{\gamma }^4})\nonumber\\
{r_0}&=&{s_0}+2\gamma^2s_1
    +O({{\gamma }^4}).
\eeq
The difference between $\gamma$ and $1/a$ enters at next-to-leading order (NLO), whereas the difference between $s_0$ and $r_0$ is only important at orders higher than next-to-next-to-leading order (NNLO).
Combining Eqns. \refeq{ERE2} and \refeq{pcot2A} gives
\beq
\mcal{A}&=&-\frac{4\pi}{M}\frac{1}{\gamma+ip}\[1+\frac{{s_0}}{2}
  \frac{{p^2}+{{\gamma }^2}}{\gamma +ip}
   +{{\(\frac{{s_0}}{2}\)}^2}
    \frac{{{({p^2}+{{\gamma }^2})}^2}}
     {{{(\gamma +ip)}^2}} \right.\nonumber\\
& &\left.\hspace{1.7 in}+\(\frac{{s_0}}{2}
  \)^3\frac{{{({p^2}+{{\gamma }^2})}^
        3}}{{{(\gamma +ip)}^3}}
    +\frac{{s_1}}{2}
     \frac{{{({p^2}+{{\gamma }^2})}^2}}
      {\gamma +ip} +\cdots \]\nonumber\\
&=&-\frac{4\pi}{M}\frac{1}{\gamma+ip}\[1+\frac{{s_0}}{2}
   (\gamma -i p) + 
   {{\(\frac{{s_0}}{2}\)}^2}
   {{(\gamma -ip)}^2} \right.\nonumber\\
& &\left.\hspace{1.7 in}+ \(\frac{{s_0}}{2}
  \)^3{{(\gamma -ip)}^3}
     +\frac{{s_1}}{2}
      {{(\gamma -ip)}^2}
       (\gamma  + ip)+\cdots \].
\eqn{EREinA}
\eeq

\lskip

The EFT for nucleons with explicit pions has already been used \cite{ksw1} to calculate the ${}^1S_0$ phase shift at LO and NLO. 
An NNLO calculation is needed primarily to check the convergence of the EFT expansion. More generally it serves as a probe for better understanding of different aspects of the theory, like the appropriate fitting procedure and the role of the renormalization group flow.

The scattering amplitude at NNLO gets contributions from instantaneous pion exchanges (``potential pions''), as in NLO, and also from retardation effects in pion exchange (``radiation pions'') and relativistic corrections. The relativistic corrections are very small compared to the potential pion contribution even at high momentum.
The radiation pion effects have been found to be highly suppressed due to Wigner's $SU(4)$ spin-isospin symmetry in the large scattering length limit in the singlet and triplet channel of N-N scattering\cite{mehen2}.

In this paper we analyze a toy model, with some realistic features, but without retardation pion effects and relativistic corrections. The EFT for nucleons using the power counting scheme introduced by Kaplan, Savage and Wise (KSW), is not completely understood yet, and it is beneficial to probe it further with an NNLO calculation in a ``cleaner'' setup. 
It should be noted that the EFT amplitude for this toy model and the one appropriate for the real ${}^1S_0$ case differ only by the small relativistic and radiation pions effects. The contribution of potential pions to the real amplitude is identical to the amplitude calculated below, in Sections \ref{NLO} and \ref{NNLO}.

\lskip

We calculate the ${}^1S_0$ phase
shift up to NNLO in a two-Yukawa toy model which is presented in Section~\ref{TOY}, with the corresponding effective Lagrangian described in Section~\ref{EFT}. A brief review of the KSW power counting is presented in Section~\ref{KSW}. Section \ref{NOPIONS} includes the calculation in the effective theory without pions, whereas the NLO and NNLO scattering amplitudes in the theory with pions are calculated in Sections \ref{NLO} and \ref{NNLO} respectively. In Subsection~\ref{matching} we derive constraints on some of the EFT couplings.
The renormalization group flow is discussed in Subsection~\ref{RG}, where the rest of the couplings at NNLO are fixed. 
 The comparison with the two-Yukawa phase shift is presented in Section~\ref{results}, followed by summary and conclusions in Section~\ref{SUMMARY}.

\end{section}

\begin{section}{The Toy Model}
\label{TOY}

In this section a theory which can be solved exactly is described. Further below we model the low energy behavior of this theory by an EFT.

We consider non-relativistic nucleons interacting via two instantaneous Yukawa potentials:

\beq
V(r)=-g_\pi\frac{e^{-m_\pi r}}{4\pi r}-g_\rho\frac{e^{-m_\rho r}}{4\pi r}
\eqn{V}
\eeq
($r$ is the distance between the nucleons).
This model has several appealing features. It is complex enough in the sense that the interactions involve two well separated scales as in the real world. In addition, the Yukawa potentials describe the instantaneous exchange of the $\pi$ and $\rho$ mesons (without the complications of retardation effects\cite{mehen2}).

The realistic ${}^1S_0$ one pion exchange amplitude
(here given in momentum space) is
\beq
\(-\frac{g_{A}^{2}}{2 f_{\pi}^{2}}\)\frac{p^2}{p^2+m_{\pi}^{2}}=-\frac{g_{A}^{2}}{2 f_{\pi}^{2}}+\frac{g_{A}^{2}}{2 f_{\pi}^{2}}\frac{m_{\pi}^{2}}{p^2+m_{\pi}^{2}},
\eqn{OPEP}
\eeq
where the first term is a contact interaction, and the second is the Yukawa part. 
The contact interaction is not included in the toy model. $g_\pi$ is chosen to reproduce the Yukawa part.
Comparing \Eq{OPEP} and \Eq{V}, we set
\beq
g_{\pi}=\frac{g_{A}^2 m_\pi^2}{2 f_{\pi}^2}.
\eeq

We take the nucleon mass to be $M=940\mathrm{\ MeV}$, as well as
$m_\pi=140\mathrm{\ MeV}, m_\rho=770\mathrm{\ MeV}, f_\pi=132\mathrm{\ MeV}$ and
$g_A=1.25$.
$g_\rho$ is tuned to give a large scattering length ($g_\rho=13.5\rightarrow a=-24.7\mathrm{\ fm}$) by numerically solving the Schroedinger equation with this potential.
The toy model is used in generating all the ${}^1S_0$ phase shifts.

\end{section}

\begin{section}{The Effective Lagrangian}
\label{EFT}

The effective Lagrangian for the toy model is described in this section. It is appropriate for center of mass energies much smaller than $m_\rho$. The effects of the short range force are reproduced in the EFT by contact interactions. As noted in \cite{ksw2}, it is convenient to separate the effective
Lagrangian in terms of the number of nucleon fields present: 
\beq {\cal L} = {\cal L}_1 + {\cal L}_2 + ...,\nonumber \eeq 
where ${\cal L}_n$ describes the propagation or scattering of
  $n$ nucleons. $\mcal{L}_1$ is the usual
 kinetic energy term\cite{ksw2,chiral} 
\beq \eqn{chiral_Lagrangian}
{\cal L}_1 = N^\dagger (i \partial_0 + \vec{\nabla}^2/2 M)N,
\eeq 
where the isodoublet field N represents the nucleons:
\beq N =\left( \begin{array}{c}
                p\\
                n
              \end{array}\right).
\nonumber 
\eeq 
Next, we write ${\cal L}_{2}^{({}^1S_0)}$, the part of ${\cal L}_2$ 
that has non-vanishing matrix elements between ${}^1S_0$ states. 
 The matrices $P_k$ \cite{ksw2} are used to project onto the ${}^1S_0$
 state,  
\beq P_k \equiv \frac{1}{\sqrt{8}}\sigma_2\otimes\tau_2\tau_k,
\nonumber 
\eeq
where the $\sigma$ matrices act on the nucleon spin space and the
$\tau$ matrices act on the nucleon isospin space. The result is\cite{ksw1,ksw2}:  

\beq
\eqn{L_2} 
{\cal L}_{2}^{({}^1S_0)} &=& -{\(\frac{\mu}{2}\)}^{4-D}(C_0 + D_2 m_\pi^2 + D_4
m_\pi^4+\cdots)(N^T P_i N)^\dagger(N^T P_i N)\nonumber\\ 
&{ }& +{\(\frac{\mu}{2}\)}^{4-D}(\frac{1}{8}C_2+ \frac{1}{8}D_4^{(2)}
  m_\pi^2+\cdots)[(N^T P_i N)^\dagger(N^T
P_i(\stackrel{\leftrightarrow}{\nabla})^2 N)+h.c]\nonumber\\ 
&{ }& + \cdots
\eeq
(summation over the repeated isospin index $i$ is implied.)

We use dimensional regularization where the Lagrangian is given in $D$ space-time 
dimensions and $\mu /2$ is an arbitrary mass scale introduced so that the
couplings $C_{2n}$, $D_{2n}^{(2m)}$ have the same units in any dimension $D$. $C_{2n}$ are the coefficients of contact interactions
containing $2n$ derivatives and
 $D_{2n}^{(2m)}$ are coefficients of contact interactions involving 
 $2m$ derivatives and $2(n-m)$ powers of the pion mass. We use  
   the convention $D_{2n}\equiv D_{2n}^{(0)},\ C_{2n}\equiv D_{2n}^{(2n)}$. The ellipses indicate
operators with higher powers of derivatives or pion mass insertions.

For $n>0$ ($m>0$), the couplings $C_{2n}$($D_{2n}^{(2m)}$) multiply a
linear combination of operators . These operators differ in the way the derivatives act on the legs
entering the vertex. In many cases factors of internal loop momentum $q$ result in factors
of external momentum $p$, and there is no need to distinguish between
the different parts that make up a $C_{2n}$($D_{2n}^{(2m)}$) vertex. At NNLO the first exception to this rule is encountered. We come back
to this point in Section~\ref{onepion}.

The interaction with the pions is described by a non-local term in the action:
\beq
S_{\pi}&=&\int d^D x\  d^D y\(N^T(x)P_i N(x)\)^\dagger g_{\pi}\frac{e^{-m_\pi|{\bf x}-{\bf y}|}}{4\pi|{\bf x}-{\bf y}|}\delta(x^0-y^0)\(N^T(y)P_i N(y)\) \nonumber\\
&=&\int d^D p\  d^D p'\(N^T(p)P_i N(p)\)^\dagger \peta\frac{m_{\pi}^{2}}{{|{\bf p}-{\bf p'}|}^2+m_{\pi}^{2}}\(N^T(p')P_i N(p')\). 
\eeq

Once the effective Lagrangian has been described, one needs a power counting scheme, i.e. a set of rules that determine the relative sizes of diagrams contributing to the scattering process.

\end{section}

\begin{section}{The Kaplan-Savage-Wise Power Counting}
\label{KSW}

The KSW power counting scheme was introduced in \cite{ksw1}, and is designed to incorporate the effects of an unnaturally large scattering length.

The two versions of the ERE, \Eq{ERE} and \Eq{ERE2}, lead to two different expansions of the amplitude: \Eq{EREinAksw} and \Eq{EREinA}.
Both expansions coincide in the infinite scattering length limit, but they are not identical. In the next section, the effective theory without pions is revisited, mostly to demonstrate some of the consequences of using  expansion \Eq{EREinA}. In this section, however, we concentrate on the KSW power counting scheme which follows from either expansion.

It has been shown \cite{weinberg} that the leading term in expansion \Eq{EREinA} is obtained by summing up the bubble chains in Fig. \ref{LOfig} with insertions of $C_0$, giving the LO amplitude:
\makefig{LOfig}{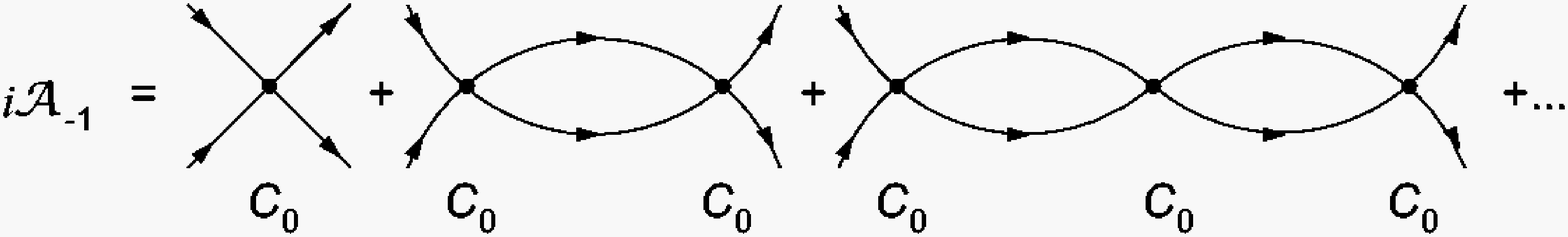}{1.1}{$\mcal{A}_{-1}$}
\beq
  \mcal{A}_{-1}= \frac{-C_0}{1 + i C_0 L},
   \eeq where $L$ is the loop integral
\beq \eqn{L}
L &=& {\( \frac \mu 2\)}^{\rm 4-D}\int\ddq \frac i {\frac E 2 -
q_0 -\frac {q^2} {2 M} + i\epsilon} \frac i {\frac E 2 + q_0
-\frac {q^2} {2 M} + i\epsilon}\nonumber\\ 
&=& -iM {\(\frac \mu
2\)}^{\rm 4-D}\int\ddqm \frac 1 { q^2 - E M
-i\epsilon}=-i\frac M{4 \pi}(X_{sub} + i p).
\eeq

A subtraction scheme that reproduces the KSW power counting is the power divergence subtraction (PDS) scheme that was introduced in
\cite{ksw1}. In PDS one subtracts the poles of $L$ in both $D=4$ and $D=3$
dimension (which gives $X_{sub}=\mu$ in \Eq{L}). Setting
$\mu\sim p$, all the diagrams contributing to the bubble chains are
rendered equal in size ($\sim 1/p$). This justifies summing up the
bubble graphs at LO.

In the theory with pions, pion effects are assumed to be perturbative.
From the PDS scheme follows a power counting
scheme - KSW - in which:

\begin{itemize}
\item{ The expansion parameter is $Q/\Lambda$, where $p,\mu,m_\pi\sim Q$.}
\item{ $\mcal{A}=\sum_{n=-1}^{\infty}\mcal{A}_n,\ \mcal{A}_n\sim Q^n$.}
\item{ The renormalized coupling $C_0=C_0(\mu)$ scales as
    ${1/\mu}\sim 1/Q$, and more generally the renormalized couplings $C_{2n}$, $D_{2n}^{(2m)}$
    scale as $ 1/{Q^{n+1}}$.}
\end{itemize}
Some immediate consequences are:
\begin{itemize}
\item{ The loop integral $L$ scales like $p$ or $\mu$, i.e. ${\cal
      O}(Q)$. Therefore, adding a $C_0$ and a loop $L$ to a diagram
    does not change its order since the powers of $Q$ cancel. It
    is now required to dress each diagram in the theory by $C_0$ to all
    orders. (This is diagrammatically represented by the ``blob'', Fig.~\ref{blobfig}).

\makefig{blobfig}{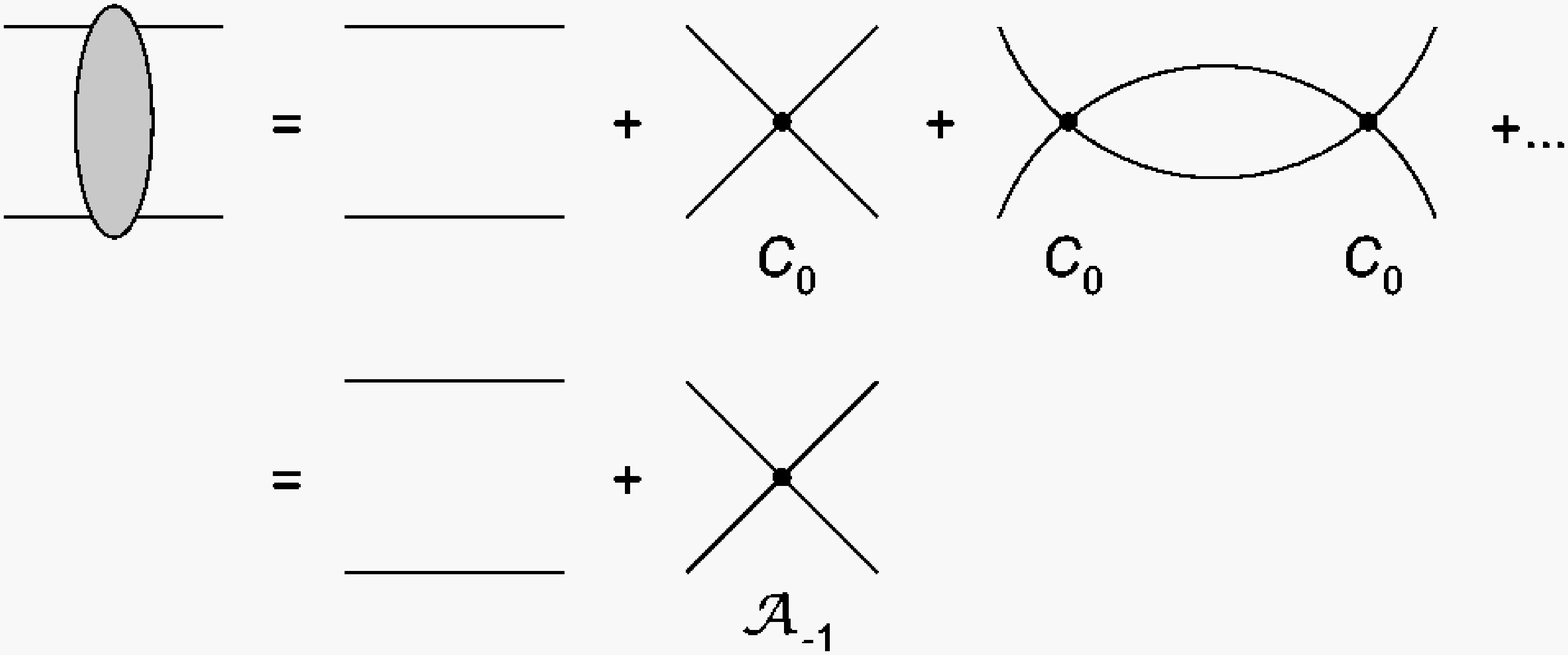}{2}{Definition of the ``blob''} }

\item{ When the nucleons are put on shell, nucleon propagators scale
    as $1/Q^2$ and loop integrals $\int d q^4$ scale as $Q^5$.}
\end{itemize}
 
This theory should correctly describe the physics at momentum below
$\Lambda\sim m_\rho/2$, which is the scale associated with the cut in the amplitude due to $\rho$ exchange. A calculation to order $Q^n$ is expected to be accurate up
to an error of $\sim(Q/\Lambda)^{n+1}$.
\lskip

In the last three sections, all the components of the computational framework of the EFT for nucleons have been identified: the effective Lagrangian (Feynman rules), the regularization and subtraction methods, and the power counting scheme that organizes the diagrams of the theory in a perturbative series. In the following sections we apply this formalism to the calculation of the ${}^1S_0$ scattering amplitude.
\end{section}

\begin{section}{The Effective Theory Without Pions}
\label{NOPIONS}

At energies low compared to the pion mass, even the long range pion exchange is effectively point-like. Both the real N-N scattering process and the one in the two-Yukawa theory are modeled by the same EFT without pions at these energies, except for relativistic corrections. The theory without pions has been solved before \cite{ksw1}, using expansion \Eq{EREinAksw}. However, since we use the $Q$ expansion \Eq{EREinA}, it is instructive to consider the effective theory without pions again.

Expansion \Eq{EREinAksw} is not well behaved near the pole, at momentum $p\sim i/a$. This is because higher orders in the expansion contain terms more singular than the leading order term. This means that the $Q$ expansion breaks down close to the pole.
Expansion \Eq{EREinA} avoids this problem. All terms have only a simple pole and the expansion is not expected to fail even at very low $p$.

The structure of the effective Lagrangian when pion effects are integrated out can be obtained from the Lagrangian that includes pions by dropping all the $D$ couplings, as well as  the interaction part $S_{\pi}$. Note that this only gives the structure of the Lagrangian, but the meaning of the $C$ couplings is different in the two EFTs, and the relation between them can only be seen by comparing the theories in a region where they are both valid. 

Expansion \Eq{EREinA} is the exact description of the phase shift at low energies. The EFT expansion, $\mcal{A}=\mcal{A}_{-1}+\mcal{A}_0+\mcal{A}_1+\cdots$, should recover it. We equate the two expansions order by order and solve for the couplings $C_{2n}$ in terms of the parameters $\gamma,\ s_0,\ s_1,\ldots$
Matching the LO amplitude with expansion  \Eq{EREinA}, we find
\beq
C_0=-\frac{4 \pi}{M}\left ( \frac{1}{\mu -\gamma} \right).
 \eeq
At NLO, dressing the $C_2p^2$ operators with $C_0$ bubbles yields 
\beq
\mcal A_0=-C_2 p^2 \left (\frac{-\mcal A_{-1}}{C_0}\right)^2\sim\frac{p^2}{(\gamma+ip)^2}. 
\nonumber \eeq
This does not have the right form to reproduce the $Q$ expansion \refeq{EREinA}, which predicts
\beq
\mcal A_0\sim\frac{p^2+\gamma^2}{(\gamma+ip)^2}. 
\nonumber \eeq
To match to the $Q$ expansion at NLO, a contact operator similar to $C_0$ is needed. This suggests that a sub-leading piece from $C_0$ is required in order to reproduce the low energy expansion. More generally, we expand the parameters in powers of $Q$\cite{mehen}:   
\beq
C_{0}&=&C_{0,-1}+C_{0,0}+C_{0,1}+\cdots \nonumber\\
C_{2}&=&C_{2,-2}+C_{2,-1}+\cdots \nonumber\\
C_{4}&=&C_{4,-3}+C_{4,-2}+\cdots \nonumber\\
\vdots& &
 \eeq
where the second subscript denotes the scaling with powers of
$Q$.
The Feynman diagrams in this theory can be summed to all orders, and the EFT expansion can reproduce exactly the ERE\cite{newgautam}. Working up to NNLO we find in the PDS scheme
\beq
C_{2,-2} &=&\frac{M}{4 \pi}\frac{s_0}{2}C_{0,-1}^2=\frac{4 \pi}{M}\frac{s_0}{2}\left (\frac{1}{\mu-\gamma}\right )^2\nonumber\\
C_{0,0}  &=&\gamma^2 C_{2,-2}\nonumber\\
C_{4,-3} &=&\frac{C_{2,-2}^2}{C_{0,-1}}=-\frac{4 \pi}{M}\frac{s_0^2}{4}\left (\frac{1}{\mu-\gamma}\right )^3\nonumber \\
C_{2,-1} &=&2 \gamma^2 C_{4,-3}\nonumber\\
C_{0,1} &=&\gamma^4 C_{4,-3}.
\eqn{nopiCs}
\eeq

Expansions \Eq{EREinA} and \Eq{EREinAksw} coincide in the limit $a\rightarrow \infty$($\gamma\rightarrow0$). This is also demonstrated in the vanishing of the sub-leading couplings in this limit.
The couplings are given in terms of the ``experimental'' quantities $\gamma,\ s_0,\ s_1,$ etc.. At LO $C_{0,-1}$ is determined from $\gamma$, at NLO $C_{0,0}$ and $C_{2,-2}$ are given in terms of $\gamma$ and $s_0$, and at NNLO $C_{0,1},\ C_{2,-1},$ and $C_{4,-3}$ are again fixed by $\gamma$ and $s_0$. 
From \Eq{nopiCs} it can also be seen that only for the leading $C_{2n}$s, the scaling with powers of $\mu$ is the same as the scaling with powers of $Q$. This demonstrates explicitly how knowledge of the running couplings may not be enough to determine the power counting of an operator. Here, it is the expansion about the exact pole that has lead to this behavior.

To relate the scattering amplitude to phase shift, we use the relation
\beq
S=e^{2 i \delta}=1+i\frac{Mp}{2\pi}\mcal{A},
\eeq
to write
\beq \delta=\frac{1}{2i}\ln\(1+i\frac{Mp}{2\pi}\mcal{A}\),  \eeq
which can be expanded in powers of $Q$ \cite{ksw1} (writing $\delta=\delta_0+\delta_1+\cdots$):
\beq
\delta_0({C_0})&=&\frac{1}{2i}\ln\(1+i\frac{M p}{2\pi}\mcal{A}_{-1}\) \nonumber\\
\delta_1({C_0},{C_2})&=&\frac{M p}{4\pi}\frac{\mcal{A}_0}{1+i\frac{M p}{2\pi}\mcal{A}_{-1}} \nonumber\\ 
\delta_2({C_0},{C_2},C_4)&=&-i \(\frac{M p}{4\pi}\frac{\mcal{A}_0}{1+i\frac{M p}{2\pi}\mcal{A}_{-1}}\)^2+\frac{M p}{4\pi}\frac{\mcal{A}_1}{1+i\frac{M p}{2\pi}\mcal{A}_{-1}}.
\eqn{deltaexp}
\eeq

The comparison with the exact phase shift of the two-Yukawa model up to NNLO is shown in Fig.~\ref{fig:EREfit} $(a)$. 
The theory describes the ``data'' well up to $p\sim 50\mbox{\ MeV}$. To investigate the cutoff $\Lambda$ of this theory we look at the error plots\cite{error}  Fig.~\ref{fig:EREfit} $(b)$. We see that at $p\sim 200\ {\rm MeV}$, the errors from different orders in $Q$ expansion become comparable. The conditions \Eq{nopiCs} guarantee that the expansion \Eq{EREinA} is recovered. Since there are no contributions to $p\cot\delta$ at odd powers of $Q$ in this expansion, the NLO and NNLO curves in Fig.~\ref{fig:EREfit}$(b)$ are the same.

\maketwofigs{fig:EREfit}{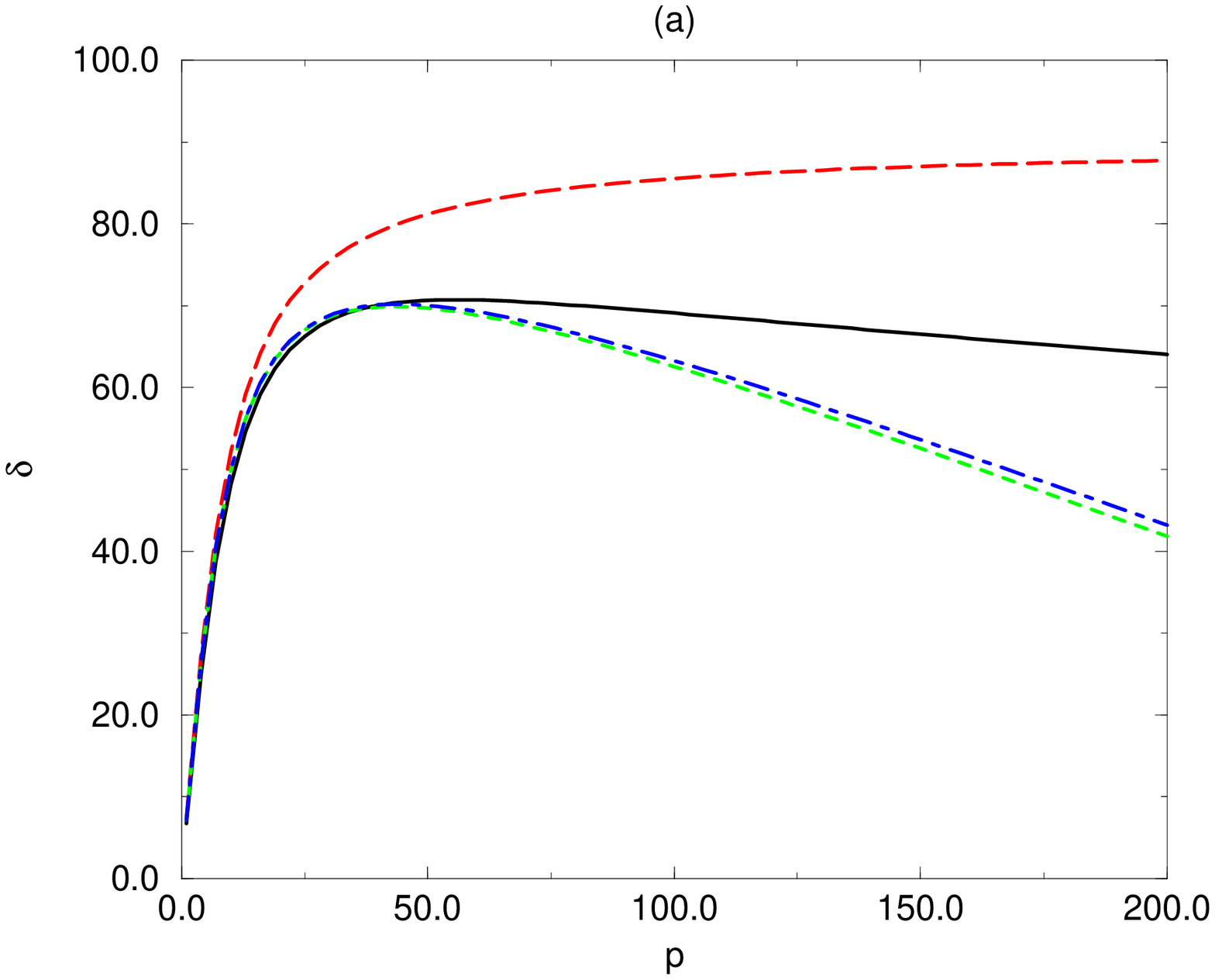}{0}{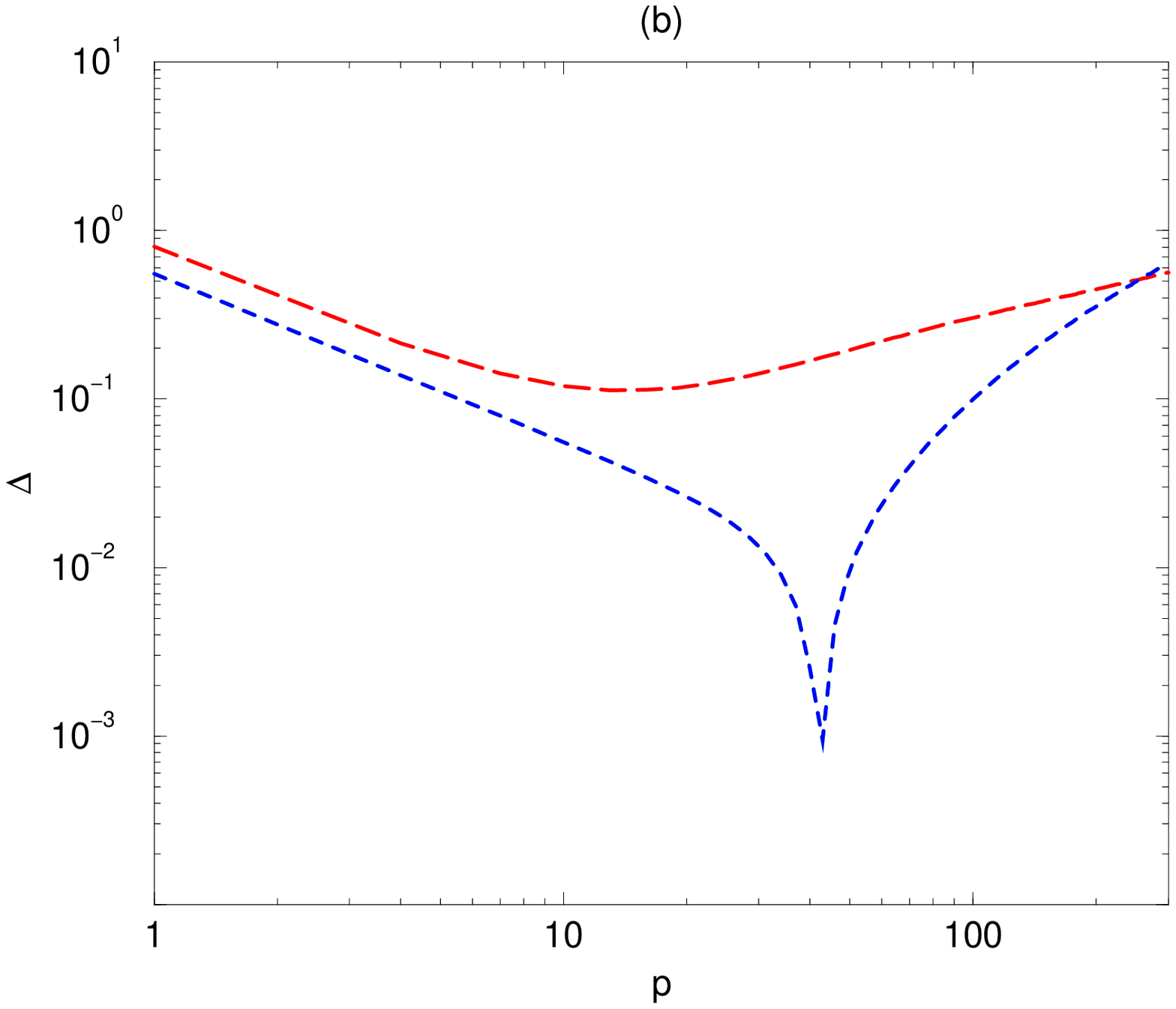}{2.9}{The Effective Theory Without Pions. (a) $\delta(degrees)$ vs. $p$(MeV) (b) log-log plot of $\Delta\equiv
  |\cot\delta_{EFT}-\cot\delta_{exact}|$ vs. $p$(MeV). The exact phase shift is described by the solid curve. The long-dashed curves denote the LO phase shift and error, the dashed curves are the NLO results, and the dot-dashed curves describe the NNLO results.}

Fig.~\ref{fig:EREfit} is produced using only the ERE. The EFT approach provides a broad formalism in which the pions are naturally included. The next two sections describe the calculation of the scattering amplitude to NLO and NNLO in the effective theory with pions.

\end{section}

\begin{section}{The Effective Theory With Pions: NLO}
\label{NLO}

 In the theory with truly dynamical pions, the contact piece of the pion exchange can be reabsorbed in the definition of $C_{0,0}$. Hence, at NLO, there is no difference in the amplitude for this toy model and the real N-N scattering amplitude as the radiation pion effects and relativistic corrections appear at NNLO. In \cite{ksw1}, the ${}^1 S_0$ phase shift has already been calculated to NLO. We include the calculation here for clarity.

At NLO, we need to consider local operators that make up the following four-nucleon vertices,
classified here according to their Q counting:\vspace{.1 in}

\indent\hspace{0.5in}\begin{tabular}{lcl}
${\cal O}(Q^{-1})$&:&$ -C_{0,-1}$\\[.1 in]
${\cal O}(Q^0)$&:&$-p^2 C_{2,-2}\ ; -C_{0,0}\ ;-m^2_\pi D_{2,-2}$\vspace{.1 in}
\end{tabular}

\noindent Only the linear combination $C_{0,0} + m^2_\pi D_{2,-2}$ appears in the amplitude, and it is denoted by  $B_{0,0}$.
From this point on we adopt the convention that the second subscript is omitted from leading couplings, so that, for instance, $C_0$ stands for $C_{0,-1}$, $C_2$ is really $C_{2,-2}$, etc. Also note that at NLO, the operator $C_2\nabla^2$ can be replaced by $C_2p^2$. As explained in Appendix B, this is not always the
  case.

The diagrams contributing to the NLO amplitude are shown in Fig.~\ref{NLOfig}. 
The amplitude is
\beq \mcal{A}_0= \mcal{A}_0^{(0 \pi)} + \mcal{A}_0^{(\pi)} \nonumber \eeq  
 where $\mcal{A}^{(0 \pi)}_0$ includes all the diagrams with no pion exchange, and $\mcal{A}^{(\pi)}_0$ contains all the diagrams with a single pion exchange.
It is convenient to express Fig.~\ref{NLOfig}(b) and Fig.~\ref{NLOfig}(c) in terms of the loop integrals $P_1$ and $P_2$, so that
\makefig{NLOfig}{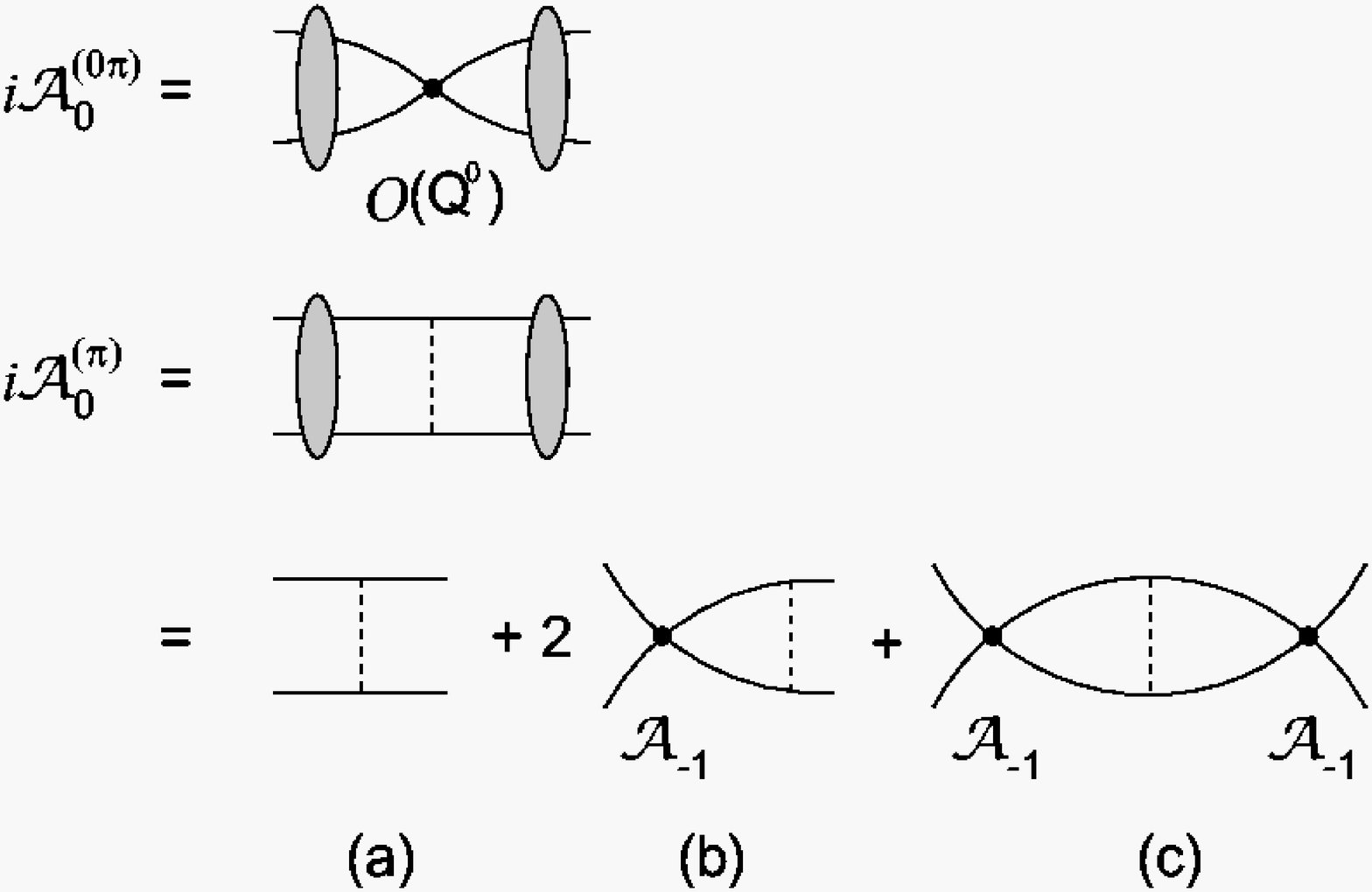}{3}{$\mcal{A}_{0}$}
\beq 
\mcal{A}_0^{(0 \pi)}&=&  {(1+ i\mcal{A}_{-1} L)}^2( - B_{0,0} - C_2 p^2)\nonumber\\
\mcal{A}_0^{(\pi)}&=& \frac{g_A^2}{2 f^2}\frac{m_\pi^2}{4 p^2} \ln (1+\frac{4
  p^2}{m_\pi^2})+ 2\ (\mcal{A}_{-1}) P_1 +i (\mcal{A}_{-1})^2 P_2, 
\eeq 
with the definitions
\beq \eqn{P1}
P_1 &\equiv&
i\peta m^2_\pi\(\frac{\mu}{2}\)^{4-D}\int\ddq\frac i {\frac E 2 + q_0 -\frac{q^2}{2 M}
+ i\epsilon}\nonumber\\
& &\hspace{2 in}\times\frac i {\frac E 2 - q_0 -\frac{q^2}{2 M} +
i\epsilon}\frac 1 {({\bf q -p})^2 + m^2_\pi} \nonumber\\ 
&=&\peta
m^2_\pi M\int\ddqm\frac 1 {q^2-{ p}^2 - i\epsilon}\frac 1
{({\bf q -p})^2+m^2_\pi}\nonumber\\
&\stackrel{PDS}\rightarrow&\frac 1 {8 \pi} \peta \frac{m^2_\pi
M}{p}\( \arctan(2 \frac p {m_\pi} ) + \frac i 2 \ln (1 +
4\frac{p^2}{m^2_\pi})\) \eeq
and
\beq \eqn{P2}
P_2
&\equiv& i\peta m^2_\pi\(\frac{\mu}{2}\)^{8-2D}\int\ddq\ddl\frac i {\frac E 2 +
q_0 -\frac{q^2}{2 M} + i\epsilon}\frac i {\frac E 2 - q_0
-\frac{q^2}{2 M} + i\epsilon}\nonumber\\
&{}&\mbox{\hspace{1 in}}\times\frac i {\frac E 2 + l_0
-\frac{{ l}^2}{2 M} + i\epsilon}\frac i {\frac E 2 - l_0
-\frac{{ l}^2}{2 M} + i\epsilon}\frac 1 {({\bf l -q})^2 +
m^2_\pi} \nonumber\\[.1 in]
&=&-i\peta m^2_\pi M^2\int\ddqm\ddlm\frac 1
{q^2-{ p}^2 - i\epsilon}\nonumber\\
& &\hspace{2 in}\times\frac 1 {{ l}^2-{ p}^2 -
i\epsilon}\frac 1 {({\bf l -q})^2+m^2_\pi}\nonumber\\
&\stackrel{PDS}\rightarrow&-i\peta \frac{m^2_\pi M^2}{32 \pi^2}
\(-\gamma_E + \ln(\pi)+1 \right.\nonumber\\
& &\hspace{1.5 in}\left.- 2\ln\(\frac{m_\pi}{\mu}\) - 2\ln\(1 -
i2\frac p {m_\pi}\) \). \eeq

\Eq{P2}, where only the poles have been subtracted, differs from an earlier calculation of $P_2$\cite{ksw1} that involved finite subtractions. The
calculation of $P_2$ in coordinate space is presented in Appendix A.

\end{section}

\begin{section}{The Effective Theory With Pions: NNLO}
\label{NNLO}

At this order, there are insertions of three kinds of local
operators, listed here
 according to their $Q$ counting:\vspace{.1 in}

\indent\hspace{0.5in}\begin{tabular}{lcl}
${\cal O}(Q^{-1})$&$:$&$ -C_0$\\[.1 in]
${\cal O}(Q^0)$&$:$&$-B_{0,0}\  ; -p^2 C_2$\\[.1 in]
${\cal O}(Q^1)$&$:$&$-B_{0,1}\  ; -p^2 B_{2,-1}\  ; -p^4 C_4$\vspace{.1 in}
\end{tabular}

\noindent There are 3 new couplings that enter at this order: $B_{0,1}\equiv C_{0,1}+m^2_\pi D_{2,-1} +m^4_\pi D_4$,  $B_{2,-1}\equiv C_{2,-1}+ m^2_\pi D_{4}^{(2)}$ and $C_4$.

The NNLO amplitude is
\beq \mcal{A}_1=\mcal{A}^{(0 \pi)}_1
+ \mcal{A}^{(\pi)}_1 + \mcal{A}^{(\pi\pi)}_1. 
\nonumber \eeq

\begin{subsection}{$\mcal{A}^{(0 \pi)}_1$}
\label{zeropion}

The diagrams in Fig.~\ref{NNLO0pfig} are generated by two insertions of ${\cal O}(Q^0)$ operators
or a single insertion of ${\cal O}(Q^1)$ operator, and give: 

\makefig{NNLO0pfig}{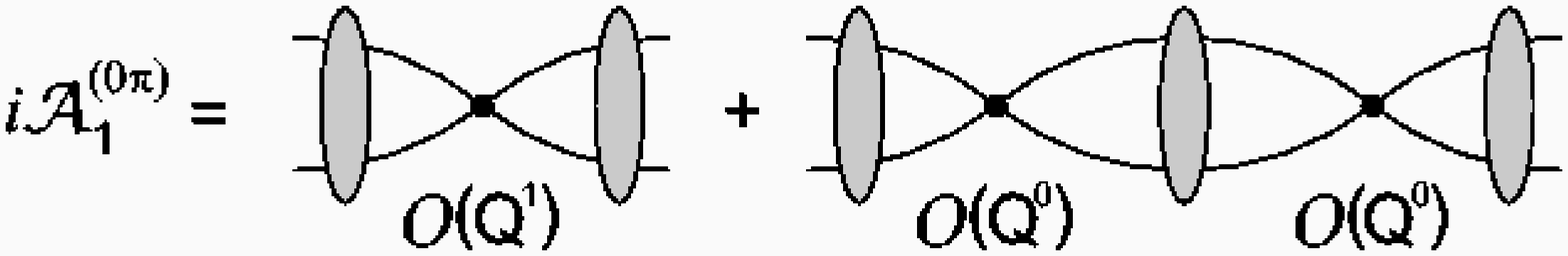}{1}{$\mcal{A}_{1}^{(0\pi)}$}

\beq
\mcal{A}^{(0 \pi)}_1 &=& {(1+ i\mcal{A}_{-1} L)}^2( - C_4 p^4 - B_{2,-1} p^2
- B_{0,1} )\nonumber\\
 &{  }& +i{(1+ i\mcal{A}_{-1} L)}^3 L {(- C_2 p^2 - B_{0,0})}^2.
 \eeq \\

\end{subsection}

\begin{subsection}{$\mcal{A}^{(\pi)}_1$}
\label{onepion}

All the diagrams of
Fig.~\ref{NNLO1pfig}  are generated by insertions of local operators
of order ${\cal O}(Q^0)$ and a single pion exchange. 

\makefig{NNLO1pfig}{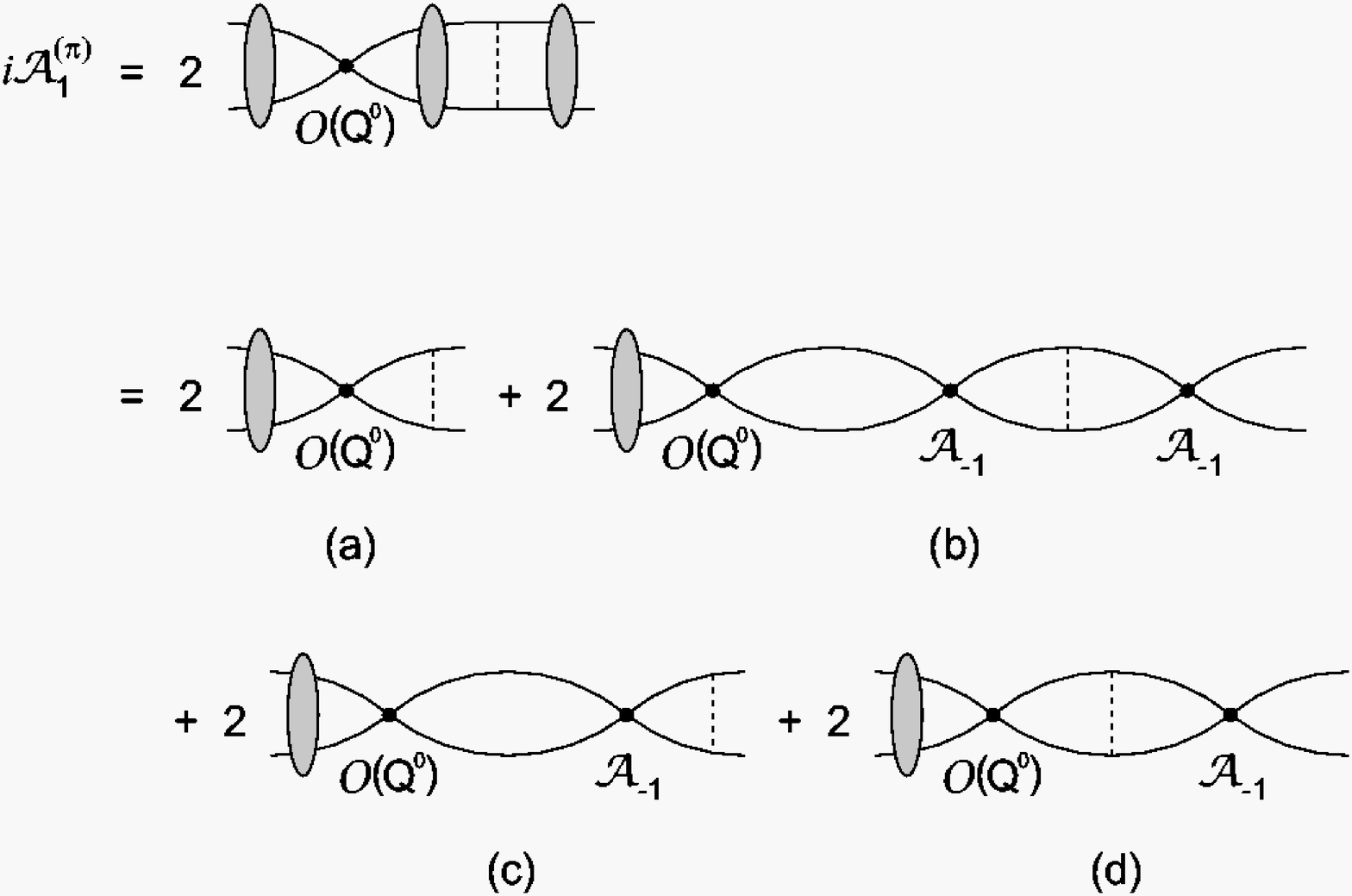}{3.5}{$\mcal{A}_{1}^{(\pi)}$}

It is here that we encounter for the first time the need to
distinguish between the operators that make up the contact vertices. 
A detailed discussion is presented in Appendix B. The result is the introduction of two new integrals, $Q_1$ and $Q_2$. The one pion exchange contribution at this order is
\beq
\mcal{A}^{(\pi)}_1
&=&2 (1 + i
\mcal{A}_{-1}L)^2(-C_2 p^2 - B_{0,0})(P_1+P_2(i
\mcal{A}_{-1}))\nonumber\\
&{  }&+2 (1 + i \mcal{A}_{-1}L)\(-\frac{C_2}{2}\)(Q_1+Q_2(i\mcal{A}_{-1})),
 \eeq
where
\beq Q_1 &\equiv&\peta m^2_\pi M\(\frac{\mu}{2}\)^{4-D}\int\ddqm\frac 1 {({\bf q- p})^2 +
m^2_\pi}\nonumber \\ &\stackrel{PDS}\rightarrow&\peta m^2_\pi M
\(\frac{\mu -m_\pi}{4 \pi}\),
 \eeq
and
\beq Q_2 &\equiv&-i\peta m^2_\pi
M^2\(\frac{\mu}{2}\)^{8-2D}\int\ddqm\ddlm\frac 1 {{ l}^2 - { p}^2 -i\epsilon}\frac 1
{({\bf q- l})^2 + m^2_\pi}\nonumber \\ &=& Q_1 L\nonumber\\
&\stackrel{PDS}\rightarrow&-i\frac{g_A^2}{2 f^2}m_\pi^2 M^2 (\frac{\mu-m_\pi}{4 \pi})(\frac{\mu + i p}{4 \pi}),  \eeq
and $P_{1,2}$ are defined in Section~\ref{NLO}.

\end{subsection}

\begin{subsection}{$\mcal{A}^{(\pi \pi)}_1$}
\label{twopion}

\makefig{NNLO2pfig}{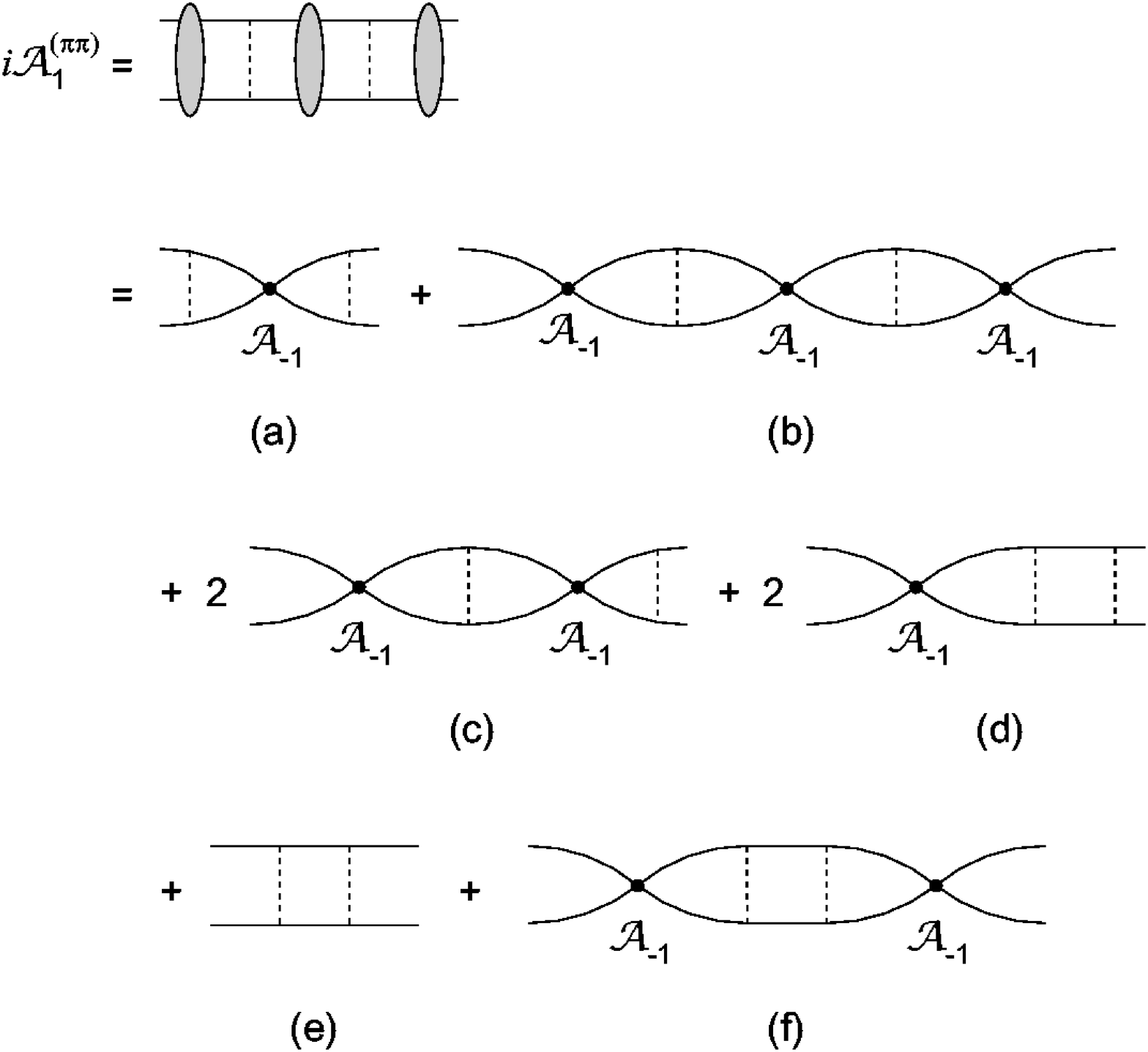}{4}{$\mcal{A}_{1}^{(\pi\pi)}$}

Fig.~\ref{NNLO2pfig} contains the diagrams of this
set. The graphs Fig.~\ref{NNLO2pfig}(e), (d) and (f) can be written in terms of
analytic functions $I_1(p)$, $I_2(p)$ and 
$I_3(p)$ respectively which are presented in some detail in 
Appendix C. The rest of the diagrams can be written in terms of
expressions that have already been
  defined. We find
\beq
\mcal{A}^{(\pi \pi)}_1
&=&-iI_1 +2(\mcal{A}_{-1})I_2
+i(\mcal{A}_{-1})^2 I_3\nonumber\\
& & + P_1(\mcal{A}_{-1})P_1 + 2 i(\mcal{A}_{-1})^2 P_1 P_2 - (\mcal{A}_{-1})^3(P_2)^2.
 \eeq

In the last two sections, we have calculated an expression for the LO+NLO+NNLO amplitude in terms of six independent free parameters $C_0$, $B_{0,0}$, $C_2$, $B_{0,1}$, $B_{2,-1}$ and $C_4$. 
When the number of parameters is doubled (from NLO to NNLO) it is hard to consider an improvement of the fit as evidence for the convergence of the EFT expansion. Therefore, it is desirable to find theoretically viable ways to reduce the number of independent parameters. This is the subject of the next section.

\end{subsection}

\end{section}

\begin{section}{Analysis}
\label{ANALYSIS}

\begin{subsection}{Matching conditions}
\label{matching}

In this subsection we introduce a method for deriving conditions that reduce the number of free parameters. The guiding principle is simple: the leading couplings are defined in such a way that certain features of the low energy phase shift (like the effective range expansion parameters) are reproduced exactly. Conditions are then imposed on the sub-leading couplings so that these features are not modified at higher orders. Similar matching conditions were introduced in \cite{mehen}.

From \Eq{pcot2A}, the
$Q$ expansion of $p\cot\delta$  in terms of the amplitude $\mcal{A}=\mcal{A}_{-1}+\mcal{A}_0+\mcal{A}_1+\cdots$, is
\beq
p\cot\delta=i p+ \frac{4\pi}{M \mcal{A}_{-1}}-\frac{4\pi}{M \mcal{A}_{-1}}\frac{\mcal{A}_0}{\mcal{A}_{-1}}+\frac{4\pi}{M \mcal{A}_{-1}}\left(\frac{\mcal{A}_0^2}{\mcal{A}_{-1}^2}-\frac{\mcal{A}_1}{\mcal{A}_{-1}}\right)+\cdots\ .
\eqn{Aexp}
\eeq
At low energies, the same quantity is given by
expansion  \refeq{ERE2}:
\beq
p\cot\delta&=&-\gamma +\frac{1}{2}s_0\left (p^2+\gamma^2\right)+\frac{1}{2}s_1\left(p^2+\gamma^2\right)^{2}+\cdots\ .
\eqn{ERE2short}
\eeq
where $p\cot\delta$ has been expanded about $p^2=-\gamma^2$ (or $p=\pm i\gamma$).

The LO amplitude contributes only a constant to $p\cot\delta$. $C_0$ is determined by requiring
\beq
i p+ \frac{4\pi}{M \mcal{A}_{-1}}=-\gamma.
\eqn{C0eqn}
\eeq
This puts the pole in the amplitude at $p=i\gamma$.
At NLO and NNLO there are more constant contributions to $p\cot\delta$. The requirements that the position of the pole does not shift at NLO, nor at NNLO, reduce to:
\beq
{\rm NLO}:\hspace{.05in} \mcal A_0|_{p=-i\gamma}&=& 0\nonumber\\
{\rm NNLO}:\hspace{.05in} \mcal A_1|_{p=-i\gamma}&=& 0.
\eqn{nogammashift}
\eeq
These conditions determine the couplings $B_{0,0}$ and $B_{0,1}$. 
Similarly, one requires that $s_0$ in Eq.~\refeq{ERE2short} is determined at NLO by $C_2$:
\beq
\frac{\partial}{\partial p^2}\(-\frac{4\pi}{M \mcal{A}_{-1}}\frac{\mcal{A}_0}{\mcal{A}_{-1}}\)|_{p^2=-\gamma^2}=\frac{1}{2}s_0,
 \eeq
and that higher order (NNLO) effects do not change $s_0$:
\beq
{\rm NNLO}:\hspace{.05in} \frac{\partial}{\partial p} \mcal A_1|_{p=-i\gamma}&=& 0.
\eqn{nos0shift}
\eeq
This condition determines $B_{2,-1}$. (Eqns.~\refeq{nogammashift} and \refeq{nos0shift} can be more directly obtained by considering the expansion of the amplitude \Eq{EREinA}.) 

Note that at NNLO, fixing $\gamma$ is not exactly the same as fitting the scattering length $a$ whereas 
reproducing $s_0$ {\em is} essentially identical to fixing the effective range, $r_0$.

Implementing the matching conditions Eqns \refeq{nogammashift} and \refeq{nos0shift}, introduces a non-analytic $m_\pi$ dependence to $B_{0,0}$, $B_{0,1}$ and $B_{2,-1}$, e.g.
\beq
B_{0,0}&=&C_2 \(\frac{4\pi }{{C_0}M}
  + \mu \)^2+\frac{{g^2}}{4{f^2}}
     m_{\pi }^{2}\left(\frac{{C_0}M}{4\pi }
      \)^2\(1-\gamma_E+\ln\(\pi \)\right.\nonumber\\
& &\mbox{\hspace{.8 in}}\left.-2 \ln\(1+
        \frac{2}{m_{\pi }}\(\frac{4\pi }{{C_0}M}+
             \mu \)
        \)+2 \ln\(\frac{\mu }{{m_{\pi }}}\)\right).
\eqn{Z1}
\eeq 
Further discussion of this point is presented in the next subsection.

The values of $\gamma$ and $s_0$ are used to determine $C_0$ and $C_2$. 
One can determine $C_4$ from $s_1$, the next free parameter in the expansion \Eq{ERE2short}. However, as in the theory without pions (\Eq{nopiCs}), it can be shown, using RG analysis, that at NNLO, $C_4$ is completely determined. This is done below.

\end{subsection}

\begin{subsection}{Renormalization Group Flow}
\label{RG}

In this subsection, RG flow equations are analyzed. We discuss how constants with $Q$ scaling, namely $\gamma$ and $m_\pi$, enter the couplings when the theory with pions is made to reproduce expansion \Eq{ERE2}.

 The RG equations are derived by requiring that the
amplitude be exactly $\mu$ independent at each order in $Q$ for any
value of $p$.
The solutions of the RG equations for $C_{0}$ and $C_{2}$ are:
\beq \eqn{C0C2D2}
 C_{0}(\mu) &=& \frac{4\pi}{M}\(\frac{1}{\xi_{C_0}}-\mu\)^{-1}\nonumber\\
C_{2}(\mu) &=& \xi_{C_2} \(\frac{M C_{0}(\mu)}{4\pi}\)^2.
\eeq 
The integration constants $\xi_{C_0}$ and $\xi_{C_2}$  are determined by
boundary conditions.
The other coupling constants, denoted here by $X$, have the general solution
\beq
X(\mu)=\xi_X \(\frac{M C_{0}(\mu)}{4\pi}\)^2+f_{X}(C_{0}(\mu),B_{0,0}(\mu),C_{2}(\mu);\mu)
\eeq
where $f_X$ is known, and the only freedom in $X$ is in the constant $\xi_X$,
determined by boundary conditions.
Specifically,
\beq
B_{0,0}&=&\left(\frac{M C_{0}}{4 \pi}\right)^2 \left(\xi_{B_{0,0}}+\peta m_{\pi}^{2}\ln\left(\frac{\mu}{M}\right)\right)\nonumber\\
B_{0,1}&=&\frac{B_{0,0}^2}{C_{0}}-\peta m_\pi^2 \frac{M\mu}{4\pi}C_{2}+\xi_{B_{0,1}}\left(\frac{M C_{0}}{4\pi}\right)^2\nonumber\\
B_{2,-1}&=&\frac{2 B_{0,0} C_{2}}{C_{0}}+\xi_{B_{2,-1}}\left(\frac{M C_{0}}{4\pi}\right)^2\nonumber\\
C_4&=&\frac{C_{2}^{2}}{C_{0}}+\xi_{C_4}\left(\frac{M C_{0}}{4\pi}\right)^2.
\eqn{constraints}
\eeq

Examination of the solutions in \Eq{constraints} shows that the $\mu$ scaling of some couplings does not agree with their expected $Q$ counting. To determine the $Q$ counting one also needs to know the $Q$ counting of the RG constants $\xi_X$.
One way of examining this without any further assumptions is by finding what values of all six parameters give reasonable fits to the data. Following this procedure we have found that at least some of the integration constants $\xi_X$ must have non-trivial $Q$ counting in order to reproduce the low energy phase shift (note that for determining the $Q$ counting, the factors of $m_\pi$ in the definitions of the $B$ couplings must be taken into account).
This result is independent of the specific form of the low energy expansion used or the particular matching conditions imposed.
Using a specific fitting procedure, we have found explicit examples of $\gamma$ and $m_\pi$ entering the couplings. 
That $\gamma$ can make up $Q$ counting was already shown in the theory without pions, where, for example, $C_{0,0}\sim \gamma^2/(\mu-\gamma)^2$ (see Eqns.~\refeq{nopiCs}). 
 
A  remark about the $m_\pi$ dependence of the theory is appropriate.  In the usual approach to EFT, the effects encapsulated in the contact interactions are the result of integrating out fields associated with heavy particles. In the case of the two Yukawa toy model, this would mean that all the couplings - $C$s and $D$s - depend only on the details of the physics at the scale $m_\rho$ where the pion is practically massless, and one would therefore expect them to be independent of $m_\pi$. More specifically, one expects $C_0$ to be a function only of the pole in the amplitude of the high energy theory, that does not include the pions, $C_0=C_0(\gamma_\rho)$.

It is a key feature of the KSW power counting scheme, however,  that at LO the amplitude has the right pole, or the correct scattering length. This implies $C_{0,-1}=C_{0,-1}(\gamma)$. Note that in general the exact pole in the amplitude has a complicated dependence on pion physics, and in particular $\gamma=\gamma(m_\pi)$. At higher orders there are contributions from pion graphs that move the position of the pole in the amplitude, but these effects have already been accounted for. Counterterms must be introduced to cancel these effects, which are in general complicated (non-analytic) functions of $m_\pi$. This has already been shown to happen: satisfying the matching conditions \Eq{nogammashift} (introduced indeed to keep the pole from moving), implied explicit $m_\pi$ dependence of $B_{0,0}$ (\Eq{Z1}).

Once it is understood that RG constants such as $m_\pi$ and $\gamma$ can enter the couplings, the $\mu$ dependence is not enough for determining the $Q$ counting. It is possible, however, that a more careful statement is true, concerning the leading couplings. We demonstrate this point by discussing directly the coupling $C_4$, which has power counting of $1/Q^3$. The second term in the solution of the RG equation for $C_4$ (\Eq{constraints}) is $\sim \xi_{C_4}/\mu^2$. In order for it to contribute at NNLO, $\xi_{C_4}$ should be $\mcal{O}(1/Q)$. Since $C_4$ has a sensible large scattering length ($\gamma\rightarrow 0$) limit, the possibility $\xi_{C_4}\sim 1/\gamma$ is ruled out. It is also assumed that $C_4$ is fundamentally independent of pion physics, and that absorbing pion effects in it (as in going from $C_0(\gamma_\rho)$ to $C_{0,-1}(\gamma)$) introduces explicit $m_\pi$ dependence only in the sub-leading couplings. These considerations allow setting $\xi_{C_4}=0$ at this order, and have $C_4$ be completely determined by $C_0$ and $C_2$.

There are in fact no new parameters at NNLO!

In the next section we check how well the EFT reproduces the phase shift of the toy model when the matching conditions derived in the last two subsections are imposed.

\end{subsection}
\end{section}

\begin{section}{Results}
\label{results}

In this section the EFT expansion is compared to the exact two Yukawa theory. We consider how well the ``data'' is reproduced by the EFT, whether the expansion is converging, and what is the breakdown scale.
The results are presented in
Fig.~\ref{fig:Fit1}(a). Fig.~\ref{fig:Fit1}(b) shows the normalized errors at each
order, $\Delta\equiv |\cot\delta_{EFT}-\cot\delta_{exact}|$ on a log-log scale.
${C_0}$ is fixed at LO by $\gamma$, and ${C_2}$ at NLO by $\gamma$ and $s_0$. By the RG equations, $C_4=C_2^2/C_0$. 

\maketwofigs{fig:Fit1}{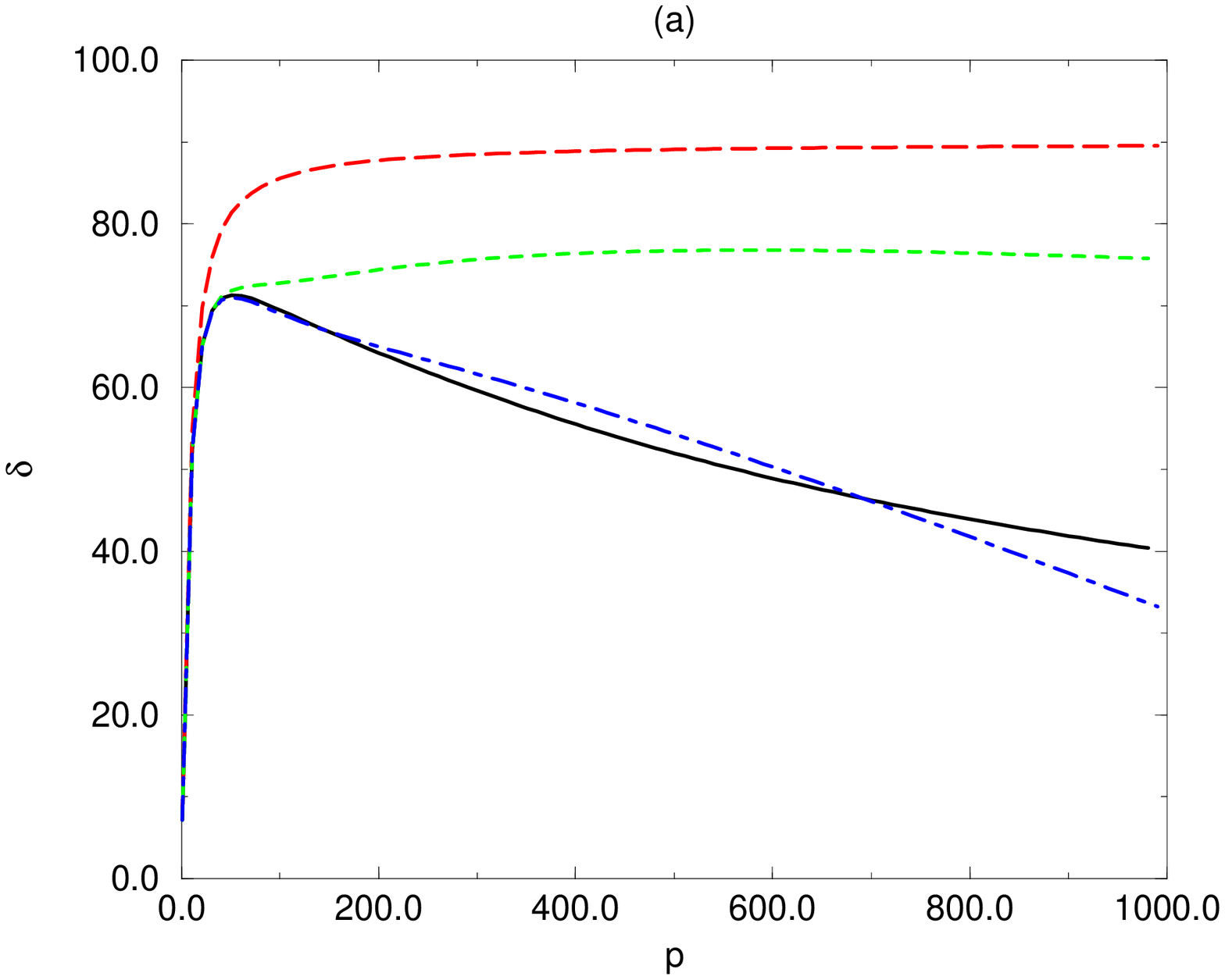}{0}{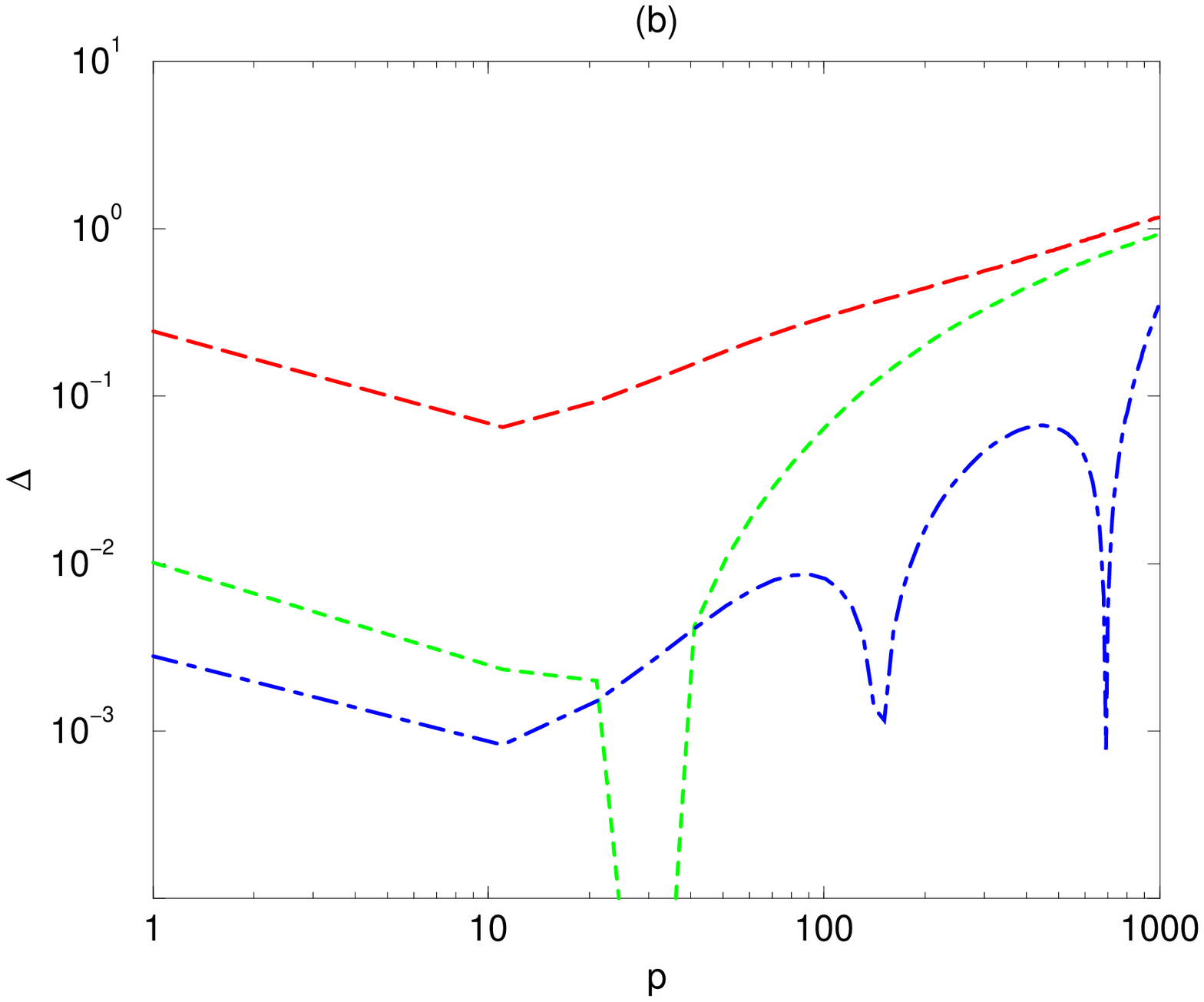}{2.4}{(a) The phase shift $\delta$ for the ${}^1S_0$ channel. (b) Log-log plot of $\Delta\equiv
  |\cot\delta_{EFT}-\cot\delta_{exact}|$ vs. $p\ $(MeV). The exact phase shift is described by the solid curve. The long-dashed curves denote the LO phase shift and error, the dashed curves are the NLO results, and the dot-dashed curves describe NNLO results.}

The graphs show a clear improvement of the fit at NLO compared to LO. There is also improvement at NNLO, without introducing any new free parameters. The error plot shows
that the errors at NNLO are smaller than those at NLO.
The breakdown scale, where the errors become comparable, can be read
off of the error plot to be $\sim 1000$ MeV. This is clearly an unrealistic bound since  $\rho$ exchange introduces a cut in the amplitude already at $m_\rho/2$.

The table below shows the couplings at $\mu=m_\pi$. The entries are given
in units of ${\rm fm}^2$. To make a direct comparison of the
  couplings easier, we included the momentum factors from the operators
  $C_2 p^2,\ B_{2,-1} p^2$ and $C_4 p^4$, evaluated with $p=m_\pi$.\vspace{.2 in}
\begin{center} 
\begin{tabular}{|c||c|c||c|c|c|}\hline
\multicolumn{1}{|c||}{LO}&\multicolumn{2}{|c||}{NLO}&\multicolumn{3}{|c|}{NNLO}\\ \hline
  $C_{0}$ & $C_{2} m_\pi^2$ & $B_{0,0}$ & $B_{0,1}$ & $\ B_{2,-1}\ m_\pi^2\ $ & $C_4\ m_\pi^4$\\ 
\hline $\  -3.526\ $ & $\ 0.06912\ $ &$\ 1.412\ $ & $\ 0.5620\ $ & $\ 0.2886\ $ & $\ -0.001355\ $\\
\hline
\end{tabular}
\end{center}
\lskip

From the table we see a hierarchy in the sizes of the couplings corresponding to the
orders in the expansion. This is a verification of the power counting scheme. A rough estimate of the expansion parameter is $Q\sim m_\pi/(m_\rho/2)\sim1/3$, which is also suggested by the table.

\end{section}

\begin{section}{Summary and Conclusions}
\label{SUMMARY}

In this paper, an NNLO calculation for the ${}^1S_0$ channel amplitude of N-N scattering for a two Yukawa model was presented. The amplitude is identical to the contribution of potential pions to the real ${}^1S_0$ scattering amplitude. 
Matching conditions for the amplitude at low momentum were derived. These matching conditions and RG constraints were used so that no new free parameters are introduced at NNLO.

We find significant improvement at NNLO over the NLO result. From the hierarchy in the sizes of the couplings we conclude that the expansion is converging. The scale where the errors from different orders in the expansion become comparable is about $1000$ MeV, which is higher than the expected breakdown scale for this theory.

The matching procedure employed in this paper could be applied to the experimental N-N scattering data. Since the relativistic corrections and radiation pion effects are small, we expect the analysis and observations in this paper to hold even for real nucleons.

\end{section}

\begin{subsection}*{Acknowledgments}

We thank David Kaplan and Martin Savage for many helpful discussions. We would also like to thank the following people for providing us with interesting and useful feedback on various subjects: Paulo Bedaque, Harald Grie{\ss}hammer, Dan Phillips and James Steele. 
This work is supported in part by the U.S. Dept. of Energy under Grants No. DE-FG03-97ER4014 and DOE-ER-40561.

\end{subsection}

\begin{section}*{Appendices}

In many cases it is easier to calculate the integrals in this theory in coordinate space. Finite integrals of this type are $I_1$, $I_2$ and $I_3$. 
When the integrals are divergent in $D=3$ and/or $D=4$, some care is needed\footnote{We would like to thank Prof. L.S. Brown for showing us how to extend the integrals off $D=4$ dimension, and specifically for providing us with an example the calculation of $P_2$ is based upon.}. $P_2$ is such an example.

The calculation of $P_2$ in coordinate space is presented in Appendix A. In Appendix B we discuss the special integrals appearing first at NNLO when a $C_2 p^2$ acts inside a loop with a pion across. $I_3$ is calculated in Appendix C, whereas for $I_1$ and $I_2$, only the result is presented.

\begin{subsection}*{Appendix A: $P_2$}
After performing the $q^0$ integral and for convenience making the integral dimensionless, \Eq{P2} gives
\beq
P_2&=&-i\peta m_\pi^{2D-6}\left(\frac{\mu}{2}\right)^{8-2D} M^2\int\ddqm\ddlm\frac{1}{q^2-\rho^2-i\epsilon}\nonumber\\
& &\hspace{2.8 in}\times\frac{1}{l^2-\rho^2-i\epsilon}\ \frac{1}{( \bfl-\bfq )^2+1},\nonumber\\
 \eeq
where $\rho = p/m_\pi$ is dimensionless. Fourier transforming to coordinate space yields
\beq
P_2&=&-i\peta m_\pi^{2 D-6}\left(\frac{\mu}{2}\right)^{8-2D} M^2\int d^{D-1} {\bf x} G_N(x) G_N(-x) G_\pi(-x),
\eeq
where 
\beq
G_N(x)&=&\int\ddkm \frac{e^{i \bfk\cdot {\bf x}}}{k^2 -\rho^2
-i\epsilon}\nonumber\\
G_\pi(x)&=&\int\ddkm \frac{e^{i \bfk\cdot {\bf x}}}{k^2 +1}
 \eeq
are the rescaled ($\rho=p/m_\pi$) Fourier transformed nucleon and pion propagators in $D$ dimensions respectively. 
Now, consider the general form
\beq I&=&\int d^{D-1}{\bf x} \prod_i G(\kappa_i ; x)\nonumber\\
&=&\frac{2 \pi^{(D-1)/2}}{\Gamma (\frac{D-1}{2})}\frac{1}{\left (2 \pi\right)^{3 (D-1)/2} }\int dx x^{D-2} \prod_i\left ( \frac{\kappa_i}{x}\right)^{\frac{D-3}{2}} K_{\frac{D-3}{2}}\left(\kappa_i x\right)\nonumber\\
&=&\frac{2 \pi^{(D-1)/2}}{\Gamma (\frac{D-1}{2})}\frac{1}{\left (2 \pi\right)^{3 (D-1)/2} }\int dx x^{7-2D} \prod_i\left ( \kappa_ix\right)^{\frac{D-3}{2}} K_{\frac{D-3}{2}}\left(\kappa_i x\right),
 \eeq
where 
\beq
G(\kappa_i ; x)&=&\int\ddkm \frac{e^{i\bfk\cdot{\bf x}}}{k^2+\kappa_i^2}\nonumber\\
&=&\frac{1}{\left (2 \pi\right)^{(D-1)/2} }\left ( \frac{\kappa_i}{ x}\right)^{\frac{D-3}{2}} K_{\frac{D-3}{2}}\left(\kappa_i x\right)\nonumber\\
&\stackrel{D\rightarrow 4}{\rightarrow}&\frac{e^{-\kappa_i x}}{4 \pi x},
\eqn{Gx} \eeq
and $K_\nu(z)$ is the modified Bessel function.
The divergence in the above integral comes from small distance x, corresponding to UV divergence. The integral is split into its finite and divergent pieces as
\beq
I&=&I^{<}+I^{>}\nonumber\\
I^{<}&=&\frac{2 \pi^{(D-1)/2}}{\Gamma (\frac{D-1}{2})}\frac{1}{\left (2 \pi\right)^{3 (D-1)/2} }\int\limits_0^\delta dx x^{7-2D} \prod_i\left ( \kappa_ix\right)^{\frac{D-3}{2}} K_{\frac{D-3}{2}}\left(\kappa_i x\right)\nonumber\\
I^{>}&=&\frac{2 \pi^{(D-1)/2}}{\Gamma (\frac{D-1}{2})}\frac{1}{\left (2 \pi\right)^{3 (D-1)/2} }\int\limits_\delta^\infty dx x^{7-2D} \prod_i\left ( \kappa_ix\right)^{\frac{D-3}{2}} K_{\frac{D-3}{2}}\left(\kappa_i x\right).
\eeq
Using the asymptotic form of the modified Bessel function
\beq
\lim\limits_{z\rightarrow 0}K_\nu(z)= \frac{2^{\nu-1}\Gamma(\nu)}{z^\nu},
\nonumber\eeq
we get for $I^<$
\beq
I^{<}&=&\frac{2 \pi^{(D-1)/2}}{\Gamma (\frac{D-1}{2})}\frac{1}{\left (2 \pi\right)^{3 (D-1)/2} } \frac{\delta^{8-2D}}{8-2D} \prod_i 2^\frac{D-5}{2}\Gamma(\frac{D-1}{2})\nonumber\\
&\stackrel{D\rightarrow4}{\longrightarrow}&-\frac{1}{32 \pi^2(D-4)}+ \frac{1+\gamma_E +\ln(4\pi)+ 2
 \ln(\delta)}{32\pi^2}+O((D-4)^1),
\eeq
The other integral $I^{>}$ is finite in $D=4$, and gives
\beq
I^{>}&=&\frac{1}{16 \pi^2}\int\limits_\delta^\infty dx x^2 \frac{e^{-(\kappa_1+\kappa_2+\kappa_3) }}{x^3}\nonumber\\
&=&-\frac{ \gamma_E +\ln(\kappa_1+\kappa_2+\kappa_3)+\ln (\delta) }{16\pi^2}
\eeq
and so,
\beq
I=I^{<}+I^{>}=-\frac{1}{32 \pi^2(D-4)}+\frac{1-\gamma_E +\ln(4\pi)- 2
 \ln(\kappa_1+\kappa_2+\kappa_3)}{32\pi^2}.
 \eeq
For the integral $P_2$, using
\beq
\kappa_1=\kappa_2=\sqrt{-\rho^2-i\epsilon}=-i\rho;\hspace{.2in}\kappa_3=1,\nonumber
\eeq
we finally have, after subtracting the pole at $D=4$,
\beq
P_2 =-i\peta m_\pi^2 M^2\frac{1}{32 \pi^2}\left( -\gamma_E +1+\ln(\pi)-2\ln(\frac{m_\pi}{\mu})-2\ln(1-2i\rho)\right).
 \eeq

\end{subsection}

\begin{subsection}*{Appendix B: $Q_1,Q_2$ }

In this appendix we discuss the action of a $C_2 p^2$ operator in a loop containing pions.

From the Lagrangian \Eq{L_2} we rewrite the operator
corresponding to $C_2$ as:
\beq
\frac {C_2}{2} (q_1^2+q_2^2) 
\eeq
with ${\bf q_1}$, ${\bf q_2}$ the single nucleon incoming and outgoing
momenta, in the center of mass coordinates. 
At tree level $ q_1= q_2= p$,
and the contribution from each piece is $\frac{C_2}{2}p^2$. This is,
in fact, the contribution from each piece of $C_2$ even when the legs 
are not external, so long as the loops attached to the vertex have no
pions. 
In the diagrams of Fig~\ref{PQfig} , however, corresponding to the types Fig.~\ref{NNLO1pfig}(a) and (d), the situation is
different. The vertex reads $\frac{C_2}{2}(p^2+q^2)$, with $q$ being
the loop momentum. Using the fact that the loop
still 
contains a nucleon propagator we write (after performing the $q^0$
integral):
\beq \frac{ q^2}{ q^2 - { p}^2 -
i\epsilon} = 1 + { p}^2\frac{1}{ q^2-{ p}^2
-i\epsilon}.\nonumber \eeq
\makefig{PQfig}{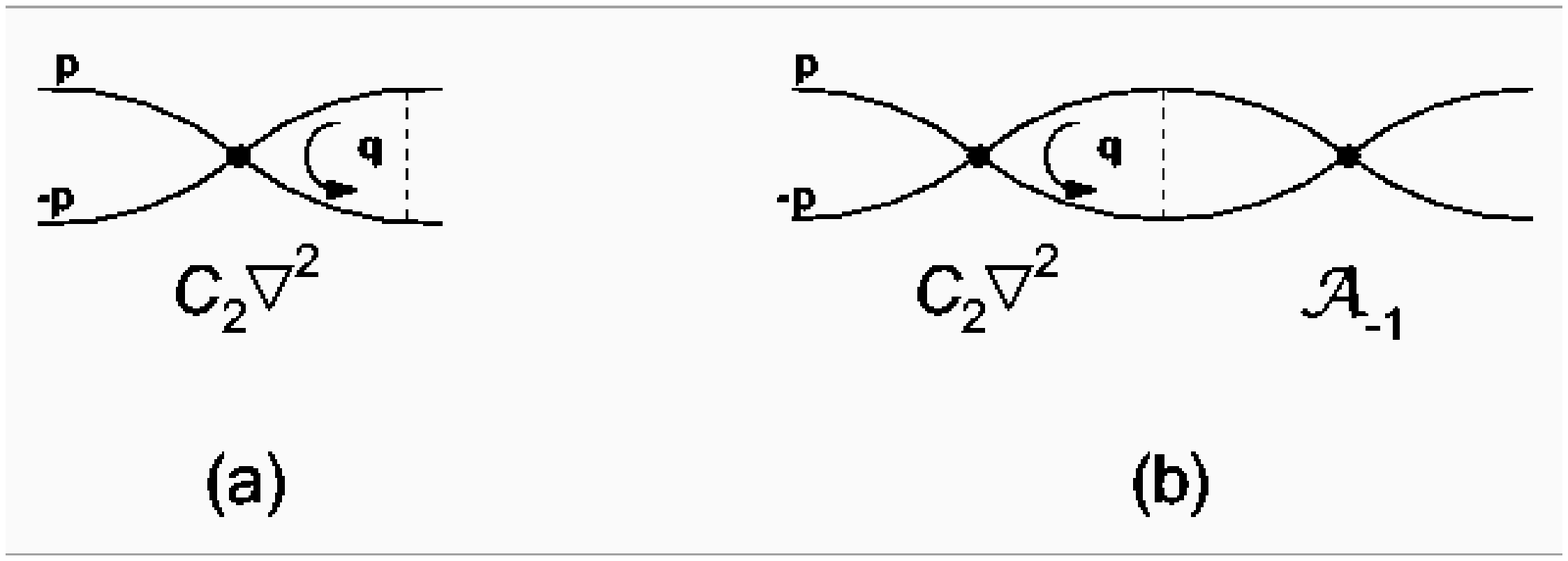}{1.8}{Diagrams in which
  one cannot use the replacement $\nabla^2\rightarrow p^2$. The
  integrals that result are (a) $P_1$ and $Q_1$, and (b) $P_2$ and $Q_2$.}
The second term looks like the $p^2$ term; the contributions (each with $C_2/2$) add up to give the
``naive'' answer (i.e. the answer one gets by replacing
$\nabla^2\rightarrow p^2$). But the ``1'' remains, giving rise to
different integrals. 

In the case of Fig.~\ref{PQfig}(a),
when the derivatives act on the external legs, the loop integral is
$P_1$ as defined in \Eq{P1}; when they act on the internal legs, the resulting integral is
\beq Q_1 &\equiv&\peta m^2_\pi M\(\frac{\mu}{2}\)^{4-D}\int\ddqm\frac 1 {({\bf q- p})^2 +
m^2_\pi}\nonumber \\ &\stackrel{PDS}\rightarrow&\peta m^2_\pi M
\(\frac{\mu -m_\pi}{4 \pi}\).
 \eeq
Similarly the contribution from diagram Fig.~\ref{PQfig}(b) can be
expressed in terms of the loop integral $P_2$ as defined in \Eq{P2} and 
\beq Q_2 &\equiv&-i\peta m^2_\pi
M^2\(\frac{\mu}{2}\)^{8-2D}\int\ddqm\ddlm\frac 1 {{ l}^2 - { p}^2 -i\epsilon}\frac 1
{({\bf q- l})^2 + m^2_\pi}\nonumber \\ &=& Q_1 L\nonumber\\
&\stackrel{PDS}\rightarrow&-i\frac{g_A^2}{2 f^2}m_\pi^2 M^2 (\frac{\mu-m_\pi}{4 \pi})(\frac{\mu + i p}{4 \pi}).  \eeq

In carrying out the subtractions in $Q_1$ and $Q_2$, some care is needed. The poles and the corresponding counterterms should be calculated treating $C_0$ perturbatively.
\end{subsection}

\begin{subsection}*{Appendix C: $I_1,I_2,I_3$ }

From Fig.~\ref{NNLO2pfig}(e), we define
\beq
I_3&=&-i \(\frac{{g^2}}{2{f^2}}\)^2 m_{\pi }^{4} {M^3} \bar{I}_3\nonumber\\
\bar{I}_3&=& \int\dtkm\frac{1}{k^2-p^2-i\epsilon}I(\bfk,\bfp)^2\nonumber\\
I(\bfk,\bfp)&=&\int\dtqm\frac{1}{q^2-p^2-i\epsilon}\frac{1}{(\bfq-\bfk)^2+m_\pi^2}.
 \eeq
$I(\bfk,\bfp)$ is finite in $D=4$ and one can therefore use the simple form of the propagators in coordinate space ,\Eq{Gx}, which gives
\beq
I(\bfk,\bfp)=\frac{1}{4\pi}\int\limits_0^\infty dx \frac{\sin(x k)}{x k} e^{-(m_\pi-ip)x}, 
\eeq
and so,
\beq
\bar{I}_3&=&\frac{1}{32\pi^4}\int\limits_0^\infty dk dx dy \frac{e^{-(m_\pi-i p)(x+y)}}{x y}\frac{\sin(x k)\sin(y k)}{k^2-p^2-i\epsilon}\nonumber\\
&=&\frac{1}{16\pi^4}\int\limits_0^\infty ds\int\limits_0^{1/2} dt\frac{e^{-(m_\pi-i p)s}}{st(1-t)}\int\limits_0^\infty dk\frac{\sin(k s t )\sin(k s (1-t))}{k^2-p^2-i\epsilon}
 \eeq
For $p\geq 0$, performing a contour integral in $k$ gives
\beq
\bar{I}_3&=&\frac{i}{64\pi^3 p}\int\limits_0^{1/2} dt\frac{1}{t(1-t)}\int\limits_0^\infty ds e^{-(m-2i p)s}\frac{e^{-i 2 p s t}-1}{s}\nonumber\\
&=& \frac{i}{64 \pi^3 \rho m_\pi}\left( \ln\left(\frac{1-i\rho}{1-i 2\rho}\right)\ln(i\rho)+Li_2\left(\frac{-i\rho}{1-i2\rho}\right)\right.\nonumber\\
&{}&+\left. Li_2(1-i\rho)-Li_2(1-i2\rho)\right)
\eeq
where $\rho =p/m_\pi$ and the dilogarithm, $Li_2(z)$, is defined as

\beq
Li_2(z)=-\int\limits_0^z dt \frac{\ln(1-t)}{t}\nonumber.
\eeq

$I_1$ and $I_2$ are determined from Fig.~\ref{NNLO2pfig} (e), (d):

\beq
I_1&=&i\(\frac{{g^2}}{2{f^2}}
  \)^2 m_{\pi }^{4} M \bar{I}_1\nonumber\\
\bar{I}_1&=&\left<\int\dtqm \frac{1}{q^2-p^2-i\epsilon}\frac{1}{(\bfq-\bfp)^2+m_\pi^2}\frac{1}{(\bfq-\bfpp)^2+m_\pi^2}\right>\nonumber\\
&=&\frac{1}{32\pi m_\pi^3 \rho^3}\left(i L(\rho)+ M(\rho)\right)\nonumber\\
L(\rho)&=&\int\limits_0^{1/2} dt \frac{1}{t(1-t)} \ln\left(\frac{(1+4\rho^2 t )(1+4\rho^2(1-t))}{1+4\rho^2}\right)\nonumber\\
&=&{\frac{1}{2}{{\ln^2 (1 + 4\,{{\rho }^2})}}}\nonumber\\
M(\rho)&=&2\int\limits_0^{1/2} dt \frac{1}{t(1-t)}\left(\tan^{-1}(2\rho\sqrt{1+4\rho^2t(1-t)})-\tan^{-1}(2\rho)\right)\nonumber\\
&=&{\frac{i}{2}}\,{{\pi }^2} - i\,{{\ln^2 \(2\)}} + 
  {\frac{i}{2}}\,\ln \(4\)\,\ln \(1 - 2\,i\,\rho \) + 
  2\,\pi \,\ln \(2 - 2\,i\,\rho \)
\nonumber\\& &\hspace{.2 in} 
- i\,\ln \(2\,{{\( 1 + i\,\rho  \) }^2}\,
     \( 1 - 2\,i\,\rho  \) \)\,
   \ln \(1 + 2\,i\,\rho \) - 
  \tan^{-1} \(2\,\rho \)\,\ln \({\frac{4\,
        {{\( 1 + {{\rho }^2} \) }^2}}{1 + 
        4\,{{\rho }^2}}}\) 
\nonumber\\& &\hspace{.4 in} 
- 
  2\,\tan^{-1} \(\rho \)\,\ln \(1 + 4\,{{\rho }^2}\) - 
  2\,i\,Li_2\(
    {\frac{1}{2}} + i\,\rho \) 
\nonumber\\& &\hspace{.6 in} 
- 
  2\,i\,Li_2\(-1 - 2\,i\,\rho \) - 
  2\,i\,Li_2\(2 - 2\,i\,\rho \),
\eeq 
where the average is over the outgoing momentum $\bfpp$,
and
\beq
I_2&=&\(\frac{{g^2}}{2{f^2}}\)^2 m_{\pi }^{4} {M^2}\bar{I}_2\nonumber\\
\bar{I}_2&=&\int\dtqm\dtlm \frac{1}{q^2-p^2-i\epsilon}\frac{1}{(\bfq-\bfp)^2+m_\pi^2}\frac{1}{l^2-p^2-i\epsilon}\frac{1}{(\bfq-\bfl)^2+m_\pi^2}\nonumber\\
&=&\frac{i}{128 \pi^3 m_{\pi}^2\rho}\int_{-\infty}^{\infty}dx{\ln\(\frac{i+x+\rho}{i-x+\rho}\)\ln\(\frac{(x+\rho)^2+1}{(x-\rho)^2+1}\)}\frac{1}{x^2-\rho^2-i\epsilon}\nonumber\\
&=&
\frac{1}{128\,
     {{\pi }^2}\,{{\rho }^2}\,{{{m_{\pi }}}^2}}
\(-{{\pi }^2} + 4\,i\,\pi \,
      \ln \(2 - 2\,i\,\rho \) + 
     \ln \(2\)\,\ln \(1 + 2\,i\,\rho \) 
\right. \nonumber \\ & &\hspace{.6 in} \left.
+ 
     \ln \(1 - 2\,i\,\rho \)\,
      \ln \({\frac{8\,{{\( 1 - i\,\rho  \) }^
             4}}{{{\( 1 - 2\,i\,\rho  \) }^2}\,
           \( 1 + 2\,i\,\rho  \) }}\) 
- 
     Li_2\({\frac{1}{2}} - i\,\rho \) 
\right. \nonumber \\ & &\hspace{.8 in} \left.
+ 
     Li_2\({\frac{1}{2}} + i\,\rho \) + 
     4\,Li_2\(2 - 2\,i\,\rho \)\).
\eeq

The explicit expressions for $I_1$, $I_2$ and $I_3$ presented above are valid only for $\rho \geq 0$ and small positive imaginary $\rho$ ($|\rho|<<1$). For other values of $\rho$, the integrals have to be done taking into account the poles and cuts in the integrand appropriately.

\end{subsection}

\end{section}



\end{document}